%% file: 13786.tex
\documentclass[structabstract]{aa}  % version paper 
\usepackage{graphicx}
\usepackage{natbib}                         
\usepackage{txfonts}
\begin{document}
   \title{Observing and modeling the dynamic atmosphere of the low mass-loss C-star \object{R\,Sculptoris} at high angular resolution\thanks{Based on observations made with the Very Large Telescope Interferometer at Paranal Observatory under programs 60.A-9220, 074.D-0601, 077.D-0294 (French Guaranteed Time Observation), 078.D-0112 (Belgian Guaranteed Time Observation), and 078.D-0122 (French Guaranteed Time Observation).}}

\titlerunning{The dynamic atmosphere of \object{R\,Scl}}

\authorrunning{S.~Sacuto et~al.} 

   \author{S.~Sacuto\inst{1}, B.~Aringer\inst{2}, J.~Hron\inst{1}, W.~Nowotny\inst{1}, C.~Paladini\inst{1}, T.~Verhoelst\inst{3}, and S.~H\"ofner\inst{4}}

   \offprints{S.~Sacuto}

\institute{University of Vienna, Department of Astronomy, T\"urkenschanzstra\ss e 17, A-1180 Vienna, Austria\\
\email{stephane.sacuto@univie.ac.at}
\and INAF-OAPD, Vicolo dell'Osservatorio 5, 35122 Padova, Italy
\and Instituut voor Sterrenkunde, K. U. Leuven, Celestijnenlaan 200D, B-3001 Leuven, Belgium
\and Department of Physics \& Astronomy, Division of Astronomy \& Space Physics, Uppsala University, Box 515, 751 20 Uppsala, Sweden
}
   \date{Received September 15, 1996}
 
  \abstract
   {We study the circumstellar environment of the carbon-rich star \object{R\,Sculptoris} using the near- and mid-infrared high spatial resolution observations from the ESO-VLTI focal instruments VINCI and MIDI, respectively.}
   {These observations aim at increasing our knowledge of the dynamic processes in play within the very close circumstellar environment where the mass loss of AGB stars is initiated.}
   {We first compare the spectro-interferometric measurements of the star at different epochs to detect the dynamic signatures of the circumstellar structures at different spatial and spectral scales. We then interpret these data using a self-consistent dynamic model atmosphere to discuss the dynamic picture deduced from the observations. Since the hydrodynamic computation needs stellar parameters as input, a considerable effort is first applied to determining these parameters.} 
   {Interferometric observations do not show any significant variability effect at the 16\,m baseline between phases 0.17 and 0.23 in the K band, and for both the 15\,m baseline between phases 0.66 and 0.97 and the 31\,m baseline between phases 0.90 and 0.97 in the N band. We find fairly good agreement between the dynamic model and the spectrophotometric data from 0.4 to 25~$\mu$m. The model agrees well with the time-dependent flux data at 8.5~$\mu$m, whereas it is too faint at 11.3 and 12.5~$\mu$m. The VINCI visibility measurements are reproduced well, meaning that the extension of the model is suitable in the K-band. In the mid-infrared, the model has the proper extension to reveal molecular structures of C$_{2}$H$_{2}$ and HCN located above the stellar photosphere. However, the windless model used is not able to reproduce the more extended and dense dusty environment.}
   {Among the different explanations for the discrepancy between the model and the measurements, the strong nonequilibrium process of dust formation is one of the most probable. The transition from windless atmospheres to models with considerable mass-loss rates occurs in a very narrow range of stellar parameters, especially for the effective temperature, the C/O ratio, and the pulsation amplitude. A denser sampling of such critical regions of the parameter space with additional models might lead to a better representation of the extended structures of low mass-loss carbon stars like \object{R\,Sculptoris}. The complete dynamic coupling of gas and dust and the approximation of grain opacities with the small-particle limit in the dynamic calculation could also contribute to the difference between the model and the data.}


   \maketitle
%
%________________________________________________________________

\section{Introduction}
\label{intro}

AGB stars are complex objects where several physical processes are in play (nucleosynthesis, convective effects, stellar pulsation, shock waves, mass-loss, molecular and dust formation, etc.). Among these mechanisms, the mass loss is particularly important for the AGB phase by limiting the maximum luminosity of the object and controlling the overall stellar evolution. The general picture of the mass-loss process of carbon-rich AGB stars is quite well understood nowadays. The pulsation of the star creates shock waves pushing the matter to conditions where both temperature and pressure allow formation of dust grains. As the opacity of amorphous carbon dust is high, grains receive enough momentum through radiative pressure to be accelerated and to drag along the gas by collisions, causing a slow outflow from the star (e.g. \citealt{fleischer92,hoefner97}). \\ 
To investigate this dynamic process, it is necessary to compare predictions of hydrodynamic atmospheric models with observations. The code described by \citet{hoefner03} solves the coupled equations of hydrodynamics, together with frequency-dependent radiative transfer, taking the time-dependent formation, growth, and evaporation of dust grains into account. Dynamic model atmospheres coming from this code have successfully reproduced line profile variations \citep{nowotny05a,nowotny05b} and time-dependent spectrometric data \citep{gautschy04} of carbon-rich stars. \\
Because of the large distance of the closest AGB stars (typically 100 to 1000 parsec), only the high angular resolution technique of long-baseline optical interferometry allows probing the regions where the mass-loss process develops. Interferometric predictions based on the dynamic atmospheric models with and without mass loss are described in detail by \citet{paladini09}. The aim of this paper is to extend the work comparing the dynamic models with near- and mid-infrared high angular resolution interferometric observations of a carbon-rich star.\\
\object{R\,Sculptoris} (\object{R\,Scl}) is a carbon-rich, semi-regular (SRa) variable star having a low mass-loss rate (1-5$\times$10$^{-7}$ M$_{\odot}$\,yr$^{-1}$: \citealt{lebertre97,gustafsson97,wong04}). The mid-infrared emission between 10 and 12~$\mu$m observed in the ISO/SWS data indicates that the star is surrounded by a warm, carbon-rich dusty shell composed of amorphous carbon (AmC) and silicon carbide (SiC) \citep{hron98}. The pulsation period of the star is 374 days \citep{whitelock97}, and the cycle-to-cycle averaged visual magnitudes vary from 6.7 to 8.1. A second period of 1804 days ($\sim$5 times the stellar pulsation period) has also been detected by \citet{whitelock97}; however, its origin is not yet understood.\\
Until now, all the interferometric studies of temporal evolution of carbon-rich stars have been based on sequences of geometric and hydrostatic models with varying stellar parameters \citep{vanbelle97,thompson02,ohnaka07}. However, as argued by \citet{hoefner03}, the only way to understand how the dynamic processes influence the atmospheric structure at different spatial scales is the use of time-dependent, self-consistent dynamic models.\\ 
In this paper, we thus present the first interpretation of spectrophotometric and interferometric data of a carbon-rich star in terms of a self-consistent dynamic model atmosphere. Near- and mid-infrared interferometric data at different baselines (15, 30, and 60\,m) will allow probes of regions ranging from the stellar photosphere to the innermost region of its dusty environment where the dynamic processes of shock waves, stellar winds, and mass loss are all at work.\\

The outline of this paper is as follows. In Sect.~\ref{obs}, we present the VINCI and MIDI observations of \object{R\,Scl} and describe the calibration performed with the corresponding data-reduction software packages developed for these instruments. Section~\ref{var} is dedicated to the variability of the star obtained from previous visual photometric and polarimetric data. We investigate the mid-infrared spectrometric and near- and mid-infrared interferometric variability of the star by comparing measurements from cycle-to-cycle and phase-to-phase. %calibrated
In Sect.~\ref{hydrodynamic_rscl}, we investigate the hydrodynamic modeling of \object{R\,Scl}. Since the hydrodynamic computation needs stellar parameters as input, a considerable effort is first applied to the determination of the corresponding stellar parameters. The last part of this section presents the results of the self-consistent hydrodynamic modeling of the spectrophotometric and of near- and mid-infrared interferometric data of the star. Finally, we conclude and give perspectives for future work in Sect.~\ref{conclu-perspec}.

\section{Observations}
\label{obs}

\begin{figure}[tbp]
\begin{center}
\includegraphics[width=8.0cm]{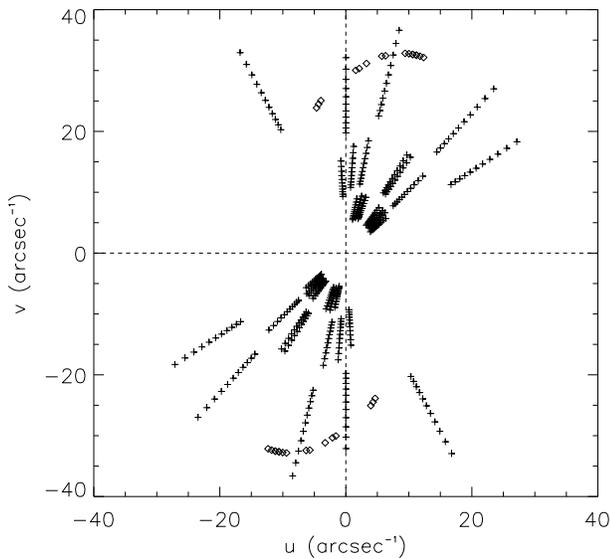}
\end{center}
\caption{K-band $uv$ coverage of the VINCI observations (diamonds) and N-band spectrally-dispersed (from 8 to 13 $\mu$m) $uv$ coverage of the MIDI observations (plus) (see Tables \ref{journal-VINCI}, \ref{journal-MIDI-UT}, and \ref{journal-MIDI-AT}).}
\label{VINCI-MIDI-uv_cov}
\end{figure}

The Very Large Telescope Interferometer (VLTI) of ESO's Paranal Observatory was used with VINCI, the near-infrared ($\lambda$=2.0-2.4 $\mu$m) interferometric recombiner \citep{kervella00}, and MIDI, the mid-infrared ($\lambda$=8.0-13 $\mu$m) interferometric recombiner \citep{leinert03}. VINCI operates around 2 microns (K band), using single spatial mode interference by employing optical fibers for spatial filtering and beam combination. MIDI combines the light of two telescopes and provides spectrally resolved visibilities in the N band atmospheric window.\\
Observations of \object{R\,Scl} were conducted in 2001, 2005, and 2006 with the VLT unit telescopes (UTs) UT2-UT4, and auxiliary telescopes (ATs) E0-G0, D0-G0, A0-G0, G0-K0, G0-H0, and D0-H0. This provides projected baselines and projected angles in the range of 11 to 64 m and 34$^{\circ}$ to 117$^{\circ}$, respectively (see Tables~\ref{journal-VINCI}, \ref{journal-MIDI-UT}, and \ref{journal-MIDI-AT}, and Fig.~\ref{VINCI-MIDI-uv_cov}). The seeing of all the VINCI observations is below 0.9$\arcsec$, whereas it is below 2.4$\arcsec$ in the case of the MIDI observations (see Tables~\ref{journal-MIDI-UT} and \ref{journal-MIDI-AT}).\\
Tables~\ref{journal-VINCI}, \ref{journal-MIDI-UT}, and \ref{journal-MIDI-AT} present the journals of interferometric observations of \object{R\,Scl} with the VINCI and MIDI focal instruments. The calibrators, \object{$\beta$\,Cet} (G9II-III, angular diameter: $\oslash$=5.18$\pm$0.06 mas), \object{$\alpha$\,CMa} (A1V, $\oslash$=5.94$\pm$0.02 mas) for VINCI and \object{HD12524} (K5III, $\oslash$=2.53$\pm$0.07 mas), \object{$\beta$\,Gru} (M5III, $\oslash$=26.8$\pm$1.3 mas), \object{$\tau^{4}$\,Eri} (M3/M4III, $\oslash$=10.6$\pm$1.0 mas), \object{HD4128} (G9II-III, $\oslash$=5.04$\pm$0.07 mas) for MIDI, were observed before or after the science target. Diameters of the VINCI calibrators were adopted from the \texttt{CHARM2} catalog \citep{richichi05}. Diameters of the MIDI calibrators \object{HD12524} and \object{HD4128} were taken from the ESO website through the \texttt{CalVin} database\footnote{\texttt{http://www.eso.org/observing/etc/}}. The diameter of the MIDI calibrator \object{$\tau^{4}$\,Eri} was taken from the \texttt{CHARM2} catalog \citep{richichi05}. Finally, the diameter of the MIDI calibrator \object{$\beta$\,Gru} was defined from the work of \citet{sacuto08}.\\ 

\begin{table}[tbp]
\caption{\label{journal-VINCI}Journal of all available VINCI observations of \object{R\,Scl}.} 
\begin{minipage}[h]{10cm}
\begin{tabular}{cccccc}\hline\hline
{\tiny Star} & {\tiny UT date \& Time} & {\tiny Phase} & {\tiny Config.} & {\tiny Base[m]} & {\tiny PA[deg]} \\
\hline
{\tiny \object{$\beta$\,Cet}} & {\tiny 2001-11-15 01:30:15} & {\tiny \ldots} & {\tiny -} & {\tiny \ldots} & {\tiny \ldots} \\
{\tiny \object{$\beta$\,Cet}} & {\tiny 2001-11-15 01:54:23} & {\tiny \ldots} & {\tiny -} & {\tiny \ldots} & {\tiny \ldots} \\
{\tiny \object{$\beta$\,Cet}} & {\tiny 2001-11-15 02:01:16} & {\tiny \ldots} & {\tiny -} & {\tiny \ldots} & {\tiny \ldots} \\
{\tiny \object{R\,Scl}} & {\tiny 2001-11-15 02:15:27} & {\tiny 0.17} & {\tiny E0-G0} & {\tiny 16} & {\tiny 69} \\
{\tiny \object{R\,Scl}} & {\tiny 2001-11-15 02:29:56} & {\tiny 0.17} & {\tiny E0-G0} & {\tiny 16} & {\tiny 71} \\
{\tiny \object{R\,Scl}} & {\tiny 2001-11-15 02:35:41} & {\tiny 0.17} & {\tiny E0-G0} & {\tiny 16} & {\tiny 71} \\
{\tiny \object{R\,Scl}} & {\tiny 2001-11-15 02:40:42} & {\tiny 0.17} & {\tiny E0-G0} & {\tiny 16} & {\tiny 72} \\
\hline
{\tiny \object{$\beta$\,Cet}} & {\tiny 2001-11-15 03:50:07} & {\tiny \ldots} & {\tiny -} & {\tiny \ldots} & {\tiny \ldots} \\
{\tiny \object{$\beta$\,Cet}} & {\tiny 2001-11-15 03:58:02} & {\tiny \ldots} & {\tiny -} & {\tiny \ldots} & {\tiny \ldots} \\
{\tiny \object{$\beta$\,Cet}} & {\tiny 2001-11-15 04:05:08} & {\tiny \ldots} & {\tiny -} & {\tiny \ldots} & {\tiny \ldots} \\
{\tiny \object{R\,Scl}} & {\tiny 2001-11-15 04:18:11} & {\tiny 0.17} & {\tiny E0-G0} & {\tiny 15} & {\tiny 84} \\
{\tiny \object{R\,Scl}} & {\tiny 2001-11-15 04:34:41} & {\tiny 0.17} & {\tiny E0-G0} & {\tiny 14} & {\tiny 86} \\
{\tiny \object{R\,Scl}} & {\tiny 2001-11-15 04:39:59} & {\tiny 0.17} & {\tiny E0-G0} & {\tiny 14} & {\tiny 87} \\
\hline
{\tiny \object{$\beta$\,Cet}} & {\tiny 2001-11-15 05:32:55} & {\tiny \ldots} & {\tiny -} & {\tiny \ldots} & {\tiny \ldots} \\
{\tiny \object{$\beta$\,Cet}} & {\tiny 2001-11-15 05:39:46} & {\tiny \ldots} & {\tiny -} & {\tiny \ldots} & {\tiny \ldots} \\
{\tiny \object{$\beta$\,Cet}} & {\tiny 2001-11-15 05:46:41} & {\tiny \ldots} & {\tiny -} & {\tiny \ldots} & {\tiny \ldots} \\
{\tiny \object{R\,Scl}} & {\tiny 2001-11-15 06:03:45} & {\tiny 0.17} & {\tiny E0-G0} & {\tiny 11} & {\tiny 99} \\
{\tiny \object{R\,Scl}} & {\tiny 2001-11-15 06:09:24} & {\tiny 0.17} & {\tiny E0-G0} & {\tiny 11} & {\tiny 100} \\
{\tiny \object{R\,Scl}} & {\tiny 2001-11-15 06:14:57} & {\tiny 0.17} & {\tiny E0-G0} & {\tiny 11} & {\tiny 101} \\
{\tiny \object{$\alpha$\,Cma}} & {\tiny 2001-11-15 06:30:06} & {\tiny \ldots} & {\tiny -} & {\tiny \ldots} & {\tiny \ldots} \\
{\tiny \object{$\alpha$\,Cma}} & {\tiny 2001-11-15 06:34:52} & {\tiny \ldots} & {\tiny -} & {\tiny \ldots} & {\tiny \ldots} \\
{\tiny \object{$\alpha$\,Cma}} & {\tiny 2001-11-15 06:46:09} & {\tiny \ldots} & {\tiny -} & {\tiny \ldots} & {\tiny \ldots} \\
{\tiny \object{$\alpha$\,Cma}} & {\tiny 2001-11-15 06:52:23} & {\tiny \ldots} & {\tiny -} & {\tiny \ldots} & {\tiny \ldots} \\
\hline
{\tiny \object{R\,Scl}} & {\tiny 2001-12-08 00:57:37} & {\tiny 0.23} & {\tiny E0-G0} & {\tiny 16} & {\tiny 70} \\
{\tiny \object{R\,Scl}$^{\dag}$} & {\tiny 2001-12-08 01:03:54} & {\tiny 0.23} & {\tiny E0-G0} & {\tiny 16} & {\tiny 72} \\ % INEXPLOITABLE (visibilite trop faible)
{\tiny \object{R\,Scl}} & {\tiny 2001-12-08 01:13:17} & {\tiny 0.23} & {\tiny E0-G0} & {\tiny 16} & {\tiny 72} \\
{\tiny \object{R\,Scl}} & {\tiny 2001-12-08 01:18:42} & {\tiny 0.23} & {\tiny E0-G0} & {\tiny 16} & {\tiny 73} \\
{\tiny \object{R\,Scl}} & {\tiny 2001-12-08 01:25:45} & {\tiny 0.23} & {\tiny E0-G0} & {\tiny 16} & {\tiny 74} \\
{\tiny \object{$\beta$\,Cet}} & {\tiny 2001-12-08 01:38:37} & {\tiny \ldots} & {\tiny -} & {\tiny \ldots} & {\tiny \ldots} \\
{\tiny \object{$\beta$\,Cet}} & {\tiny 2001-12-08 01:45:31} & {\tiny \ldots} & {\tiny -} & {\tiny \ldots} & {\tiny \ldots} \\
{\tiny \object{$\beta$\,Cet}} & {\tiny 2001-12-08 01:50:26} & {\tiny \ldots} & {\tiny -} & {\tiny \ldots} & {\tiny \ldots} \\
\hline
{\tiny \object{R\,Scl}} & {\tiny 2001-12-08 02:09:03} & {\tiny 0.23} & {\tiny E0-G0} & {\tiny 15} & {\tiny 79} \\
{\tiny \object{R\,Scl}} & {\tiny 2001-12-08 02:13:53} & {\tiny 0.23} & {\tiny E0-G0} & {\tiny 15} & {\tiny 80} \\
{\tiny \object{$\beta$\,Cet}} & {\tiny 2001-12-08 02:26:18} & {\tiny \ldots} & {\tiny -} & {\tiny \ldots} & {\tiny \ldots} \\
{\tiny \object{$\beta$\,Cet}} & {\tiny 2001-12-08 02:33:18} & {\tiny \ldots} & {\tiny -} & {\tiny \ldots} & {\tiny \ldots} \\
\hline
\end{tabular}
\end{minipage}
Note 1: The calibrators used to calibrate the visibilities are given before or after the science target. The visual phase of the star during the observations is calculated from Eq.~\ref{eq-lightcurve}. The configuration used for the observations is given. The length and position angle of the projected baseline are also indicated. Same for Tables \ref{journal-MIDI-UT} and \ref{journal-MIDI-AT}. \\
Note 2: $^{\dag}$ means that the data were not exploitable (see text). Same for Tables \ref{journal-MIDI-UT} and \ref{journal-MIDI-AT}.
\end{table}

\begin{table*}[tbp]
\caption{\label{journal-MIDI-UT}Journal of all available MIDI Unit Telescopes observations of \object{R\,Scl}.} 
\centering
\begin{tabular}{cccccccccc}\hline\hline
{\tiny \#} & {\tiny Star} & {\tiny UT date \& Time} & {\tiny Phase} & {\tiny Config.} & {\tiny Base[m]} & {\tiny PA[deg]} & {\tiny Seeing[arcsec]} & {\tiny Mode} & {\tiny Spectral resolution}\\
\hline
{\tiny 1} & {\tiny \object{R\,Scl}} & {\tiny 2005-01-03 03:01:23} & {\tiny 0.23} & {\tiny U2-U4} & {\tiny 61} & {\tiny 117} & {\tiny 1.7} & {\tiny HIGH-SENS} & {\tiny 230} \\ % P74 OUVERT
{\tiny Cal} & {\tiny \object{HD12524}} & {\tiny 2005-01-03 03:27:24} & {\tiny \ldots} & {\tiny -} & {\tiny \ldots} & {\tiny \ldots} & {\tiny 1.0} & {\tiny -} & {\tiny -}\\
\hline
{\tiny 2} & {\tiny \object{R\,Scl}$^{\dag}$} & {\tiny 2005-01-03 04:05:54} & {\tiny 0.23} & {\tiny U2-U4} & {\tiny 47} & {\tiny 136}  & {\tiny 0.9} & {\tiny HIGH-SENS} & {\tiny 230}\\ % INEXPLOITABLE : visibilites tres sup 1
{\tiny Cal} & {\tiny \object{HD12524}} & {\tiny 2005-01-03 04:45:32} & {\tiny \ldots} & {\tiny -} & {\tiny \ldots} & {\tiny \ldots}  & {\tiny 0.8} & {\tiny -} & {\tiny -}\\ % P74 OUVERT
\hline
\end{tabular}
\end{table*}

\begin{table*}[tbp]
\caption{\label{journal-MIDI-AT}Journal of all available MIDI Auxiliary Telescopes observations of \object{R\,Scl}.} %($\lambda$=8-13 $\mu$m)
\centering
\begin{tabular}{cccccccccc}\hline\hline
{\tiny \#} & {\tiny Star} & {\tiny UT date \& Time} & {\tiny Phase} & {\tiny Config.} & {\tiny Base[m]} & {\tiny PA[deg]} & {\tiny Seeing[arcsec]} & {\tiny Mode} & {\tiny Spectral resolution}\\
\hline
{\tiny 3} & {\tiny \object{R\,Scl}} & {\tiny 2006-06-16 09:38:09} & {\tiny 0.64} & {\tiny E0-G0} & {\tiny 14} & {\tiny 42} & {\tiny 0.7} & {\tiny SCI-PHOT} & {\tiny 30} \\ % P77 GTO
{\tiny Cal} & {\tiny \object{$\beta$\,Gru}} & {\tiny 2006-06-16 10:23:60} & {\tiny \ldots} & {\tiny -} & {\tiny \ldots} & {\tiny \ldots} & {\tiny 0.6} & {\tiny -} & {\tiny -}\\
\hline
{\tiny 4} & {\tiny \object{R\,Scl}} & {\tiny 2006-06-22 09:29:58} & {\tiny 0.66} & {\tiny D0-G0} & {\tiny 29} & {\tiny 46}  & {\tiny 0.8} & {\tiny SCI-PHOT} & {\tiny 30} \\ % P77 GTO
{\tiny Cal} & {\tiny \object{$\tau^{4}$\,Eri}} & {\tiny 2006-06-22 09:49:48} & {\tiny \ldots} & {\tiny -} & {\tiny \ldots} & {\tiny \ldots} & {\tiny 0.9} & {\tiny -} & {\tiny -}\\
\hline
{\tiny 5} & {\tiny \object{R\,Scl}$^{\dag}$} & {\tiny 2006-06-23 09:23:50} & {\tiny 0.66} & {\tiny E0-G0} & {\tiny 14} & {\tiny 45}  & {\tiny 1.0} & {\tiny SCI-PHOT} & {\tiny 30}\\ % INEXPLOITABLE : visibilite divergentes
{\tiny 6} & {\tiny \object{R\,Scl}} & {\tiny 2006-06-23 09:46:04} & {\tiny 0.66} & {\tiny E0-G0} & {\tiny 15} & {\tiny 50}  & {\tiny 1.4} & {\tiny SCI-PHOT} & {\tiny 30}\\ % P77 GTO
{\tiny Cal} & {\tiny \object{$\tau^{4}$\,Eri}} & {\tiny 2006-06-23 10:04:33} & {\tiny \ldots} & {\tiny -} & {\tiny \ldots} & {\tiny \ldots} & {\tiny 1.1} & {\tiny -} & {\tiny -}\\ 
\hline
{\tiny 7} & {\tiny \object{R\,Scl}} & {\tiny 2006-08-08 09:52:10} & {\tiny 0.79} & {\tiny A0-G0} & {\tiny 62} & {\tiny 77}  & {\tiny 1.1} & {\tiny SCI-PHOT} & {\tiny 30}\\ % P77 GTO
{\tiny Cal} & {\tiny \object{$\tau^{4}$\,Eri}} & {\tiny 2006-08-08 10:09:22} & {\tiny \ldots} & {\tiny -} & {\tiny \ldots} & {\tiny \ldots} & {\tiny 1.4} & {\tiny -} & {\tiny -}\\
\hline
{\tiny Cal} & {\tiny \object{HD4128}} & {\tiny 2006-09-17 04:56:51} & {\tiny \ldots} & {\tiny -} & {\tiny \ldots} & {\tiny \ldots}  & {\tiny 2.0} & {\tiny -} & {\tiny -}\\ % P78 OUVERT
{\tiny 8} & {\tiny \object{R\,Scl}$^{\dag}$} & {\tiny 2006-09-17 05:26:12} & {\tiny 0.89} & {\tiny A0-G0} & {\tiny 63} & {\tiny 63}  & {\tiny 1.8} & {\tiny HIGH-SENS} & {\tiny 30}\\ % INEXPLOITABLE : vis forte compa aux autres
\hline
{\tiny Cal} & {\tiny \object{HD4128}} & {\tiny 2006-09-17 05:51:43} & {\tiny \ldots} & {\tiny -} & {\tiny \ldots} & {\tiny \ldots}  & {\tiny 1.7} & {\tiny -} & {\tiny -}\\ 
{\tiny 9} & {\tiny \object{R\,Scl}$^{\dag}$} & {\tiny 2006-09-17 06:20:53} & {\tiny 0.89} & {\tiny A0-G0} & {\tiny 64} & {\tiny 71}  & {\tiny 2.4} & {\tiny HIGH-SENS} & {\tiny 30}\\ % P78 OUVERT
\hline
{\tiny Cal} & {\tiny \object{HD4128}} & {\tiny 2006-09-17 06:42:41} & {\tiny \ldots} & {\tiny -} & {\tiny \ldots} & {\tiny \ldots}  & {\tiny 2.0} & {\tiny -} & {\tiny -}\\
{\tiny 10} & {\tiny \object{R\,Scl}$^{\dag}$} & {\tiny 2006-09-17 07:06:01} & {\tiny 0.89} & {\tiny A0-G0} & {\tiny 63} & {\tiny 77}  & {\tiny 1.9} & {\tiny HIGH-SENS} & {\tiny 30}\\ % P78 OUVERT
\hline
{\tiny Cal} & {\tiny \object{HD4128}} & {\tiny 2006-09-18 03:13:16} & {\tiny \ldots} & {\tiny -} & {\tiny \ldots} & {\tiny \ldots}  & {\tiny 1.5} & {\tiny -} & {\tiny -}\\
{\tiny 11} & {\tiny \object{R\,Scl}$^{\dag}$} & {\tiny 2006-09-18 03:55:48} & {\tiny 0.89} & {\tiny A0-G0} & {\tiny 59} & {\tiny 48}  & {\tiny 1.6} & {\tiny HIGH-SENS} & {\tiny 30}\\ % P78 OUVERT
\hline
{\tiny 12} & {\tiny \object{R\,Scl}$^{\dag}$} & {\tiny 2006-09-18 04:19:15} & {\tiny 0.89} & {\tiny A0-G0} & {\tiny 60} & {\tiny 53}  & {\tiny 1.5} & {\tiny HIGH-SENS} & {\tiny 30}\\ % P78 OUVERT
{\tiny Cal} & {\tiny \object{HD4128}} & {\tiny 2006-09-18 04:39:39} & {\tiny \ldots} & {\tiny -} & {\tiny \ldots} & {\tiny \ldots}  & {\tiny 1.7} & {\tiny -} & {\tiny -}\\
\hline
{\tiny 13} & {\tiny \object{R\,Scl}$^{\dag}$} & {\tiny 2006-09-18 05:02:08} & {\tiny 0.89} & {\tiny A0-G0} & {\tiny 62} & {\tiny 60}  & {\tiny 1.6} & {\tiny HIGH-SENS} & {\tiny 30}\\ % P78 OUVERT
{\tiny Cal} & {\tiny \object{HD4128}} & {\tiny 2006-09-18 05:25:49} & {\tiny \ldots} & {\tiny -} & {\tiny \ldots} & {\tiny \ldots}  & {\tiny 1.6} & {\tiny -} & {\tiny -}\\
\hline
{\tiny Cal} & {\tiny \object{HD4128}} & {\tiny 2006-09-18 07:05:14} & {\tiny \ldots} & {\tiny -} & {\tiny \ldots} & {\tiny \ldots}  & {\tiny 1.5} & {\tiny -} & {\tiny -}\\
{\tiny 14} & {\tiny \object{R\,Scl}$^{\dag}$} & {\tiny 2006-09-18 07:34:30} & {\tiny 0.89} & {\tiny A0-G0} & {\tiny 61} & {\tiny 80}  & {\tiny 2.1} & {\tiny HIGH-SENS} & {\tiny 30}\\ % P78 OUVERT
\hline
{\tiny Cal} & {\tiny \object{HD4128}} & {\tiny 2006-09-19 07:07:06} & {\tiny \ldots} & {\tiny -} & {\tiny \ldots} & {\tiny \ldots}  & {\tiny 1.2} & {\tiny -} & {\tiny -}\\ % P78 OUVERT
{\tiny 15} & {\tiny \object{R\,Scl}$^{\dag}$} & {\tiny 2006-09-19 07:32:00} & {\tiny 0.90} & {\tiny D0-G0} & {\tiny 30} & {\tiny 81}  & {\tiny 1.3} & {\tiny HIGH-SENS} & {\tiny 30}\\ % INEXPLOITABLE : visibilite superieure a 1
\hline
{\tiny Cal} & {\tiny \object{HD4128}} & {\tiny 2006-09-19 07:56:01} & {\tiny \ldots} & {\tiny -} & {\tiny \ldots} & {\tiny \ldots}  & {\tiny 1.2} & {\tiny -} & {\tiny -}\\
{\tiny 16} & {\tiny \object{R\,Scl}$^{\dag}$} & {\tiny 2006-09-19 08:19:20} & {\tiny 0.90} & {\tiny D0-G0} & {\tiny 28} & {\tiny 87}  & {\tiny 1.5} & {\tiny HIGH-SENS} & {\tiny 30}\\ % P78 OUVERT
\hline
{\tiny Cal} & {\tiny \object{HD4128}} & {\tiny 2006-09-20 04:12:27} & {\tiny \ldots} & {\tiny -} & {\tiny \ldots} & {\tiny \ldots}  & {\tiny 1.3} & {\tiny -} & {\tiny -}\\
{\tiny 17} & {\tiny \object{R\,Scl}} & {\tiny 2006-09-20 04:36:30} & {\tiny 0.90} & {\tiny D0-G0} & {\tiny 31} & {\tiny 57}  & {\tiny 1.6} & {\tiny HIGH-SENS} & {\tiny 30}\\ % P78 OUVERT
\hline
{\tiny 18} & {\tiny \object{R\,Scl}} & {\tiny 2006-09-21 02:40:17} & {\tiny 0.90} & {\tiny G0-K0} & {\tiny 54} & {\tiny 34}  & {\tiny 1.7} & {\tiny HIGH-SENS} & {\tiny 30}\\ % P78 OUVERT
{\tiny Cal} & {\tiny \object{HD4128}} & {\tiny 2006-09-21 03:03:59} & {\tiny \ldots} & {\tiny -} & {\tiny \ldots} & {\tiny \ldots}  & {\tiny 1.1} & {\tiny -} & {\tiny -}\\
\hline
{\tiny 19} & {\tiny \object{R\,Scl}} & {\tiny 2006-10-16 03:05:26} & {\tiny 0.97} & {\tiny G0-H0} & {\tiny 31} & {\tiny 59}  & {\tiny 0.6} & {\tiny SCI-PHOT} & {\tiny 30}\\ % P78 GTO
{\tiny Cal} & {\tiny \object{$\tau^{4}$\,Eri}} & {\tiny 2006-10-16 03:28:25} & {\tiny \ldots} & {\tiny -} & {\tiny \ldots} & {\tiny \ldots}  & {\tiny 0.5} & {\tiny -} & {\tiny -}\\
\hline
{\tiny 20} & {\tiny \object{R\,Scl}} & {\tiny 2006-10-16 07:00:03} & {\tiny 0.97} & {\tiny D0-H0} & {\tiny 53} & {\tiny 90}  & {\tiny 0.6} & {\tiny SCI-PHOT} & {\tiny 30}\\ % P78 GTO
{\tiny Cal} & {\tiny \object{$\tau^{4}$\,Eri}} & {\tiny 2006-10-16 07:24:05} & {\tiny \ldots} & {\tiny -} & {\tiny \ldots} & {\tiny \ldots}  & {\tiny 0.5} & {\tiny -} & {\tiny -}\\
\hline
{\tiny 21} & {\tiny \object{R\,Scl}} & {\tiny 2006-10-17 02:05:02} & {\tiny 0.97} & {\tiny D0-H0} & {\tiny 59} & {\tiny 49}  & {\tiny 1.4} & {\tiny SCI-PHOT} & {\tiny 30}\\ % P78 GTO
{\tiny Cal} & {\tiny \object{$\tau^{4}$\,Eri}} & {\tiny 2006-10-17 02:26:29} & {\tiny \ldots} & {\tiny -} & {\tiny \ldots} & {\tiny \ldots}  & {\tiny 1.0} & {\tiny -} & {\tiny -}\\
\hline
{\tiny 22} & {\tiny \object{R\,Scl}} & {\tiny 2006-10-18 01:51:25} & {\tiny 0.97} & {\tiny E0-G0} & {\tiny 15} & {\tiny 47}  & {\tiny 0.8} & {\tiny HIGH-SENS} & {\tiny 30}\\ % P78 OUVERT
{\tiny Cal} & {\tiny \object{HD4128}} & {\tiny 2006-10-18 02:13:41} & {\tiny \ldots} & {\tiny -} & {\tiny \ldots} & {\tiny \ldots}  & {\tiny 0.9} & {\tiny -} & {\tiny -}\\
\hline
{\tiny 23} & {\tiny \object{R\,Scl}} & {\tiny 2006-10-18 02:34:58} & {\tiny 0.97} & {\tiny E0-G0} & {\tiny 15} & {\tiny 55}  & {\tiny 0.9} & {\tiny SCI-PHOT} & {\tiny 30}\\ % P78 GTO
{\tiny Cal} & {\tiny \object{$\tau^{4}$\,Eri}} & {\tiny 2006-10-18 02:59:24} & {\tiny \ldots} & {\tiny -} & {\tiny \ldots} & {\tiny \ldots}  & {\tiny 0.8} & {\tiny -} & {\tiny -}\\
\hline
{\tiny 24} & {\tiny \object{R\,Scl}} & {\tiny 2006-10-18 04:51:15} & {\tiny 0.97} & {\tiny E0-G0} & {\tiny 16} & {\tiny 75}  & {\tiny 1.0} & {\tiny HIGH-SENS} & {\tiny 30}\\ % P78 OUVERT
{\tiny Cal} & {\tiny \object{HD4128}} & {\tiny 2006-10-18 05:22:11} & {\tiny \ldots} & {\tiny -} & {\tiny \ldots} & {\tiny \ldots}  & {\tiny 1.1} & {\tiny -} & {\tiny -}\\
\hline
{\tiny 25} & {\tiny \object{R\,Scl}$^{\dag}$} & {\tiny 2006-10-19 03:28:15} & {\tiny 0.98} & {\tiny E0-G0} & {\tiny 16} & {\tiny 64}  & {\tiny 1.0} & {\tiny HIGH-SENS} & {\tiny 30}\\ % INEXPLOITABLE : visibilite superieure a 1
{\tiny Cal} & {\tiny \object{HD4128}} & {\tiny 2006-10-19 03:52:03} & {\tiny \ldots} & {\tiny -} & {\tiny \ldots} & {\tiny \ldots}  & {\tiny 0.4} & {\tiny -} & {\tiny -}\\ % P78 OUVERT
\hline
{\tiny 26} & {\tiny \object{R\,Scl}} & {\tiny 2006-10-19 04:15:04} & {\tiny 0.98} & {\tiny E0-G0} & {\tiny 16} & {\tiny 71}  & {\tiny 0.7} & {\tiny HIGH-SENS} & {\tiny 30}\\ % P78 OUVERT
{\tiny Cal} & {\tiny \object{HD4128}} & {\tiny 2006-10-19 04:37:45} & {\tiny \ldots} & {\tiny -} & {\tiny \ldots} & {\tiny \ldots}  & {\tiny 0.7} & {\tiny -} & {\tiny -}\\ 
\hline
{\tiny 27} & {\tiny \object{R\,Scl}} & {\tiny 2006-10-19 05:23:42} & {\tiny 0.98} & {\tiny E0-G0} & {\tiny 15} & {\tiny 79}  & {\tiny 0.5} & {\tiny SCI-PHOT} & {\tiny 30}\\ % P78 GTO
{\tiny Cal} & {\tiny \object{$\tau^{4}$\,Eri}} & {\tiny 2006-10-19 05:49:48} & {\tiny \ldots} & {\tiny -} & {\tiny \ldots} & {\tiny \ldots}  & {\tiny 0.4} & {\tiny -} & {\tiny -}\\
\hline
{\tiny Cal} & {\tiny \object{$\tau^{4}$\,Eri}} & {\tiny 2006-12-16 02:57:16} & {\tiny \ldots} & {\tiny -} & {\tiny \ldots} & {\tiny \ldots}  & {\tiny 1.7} & {\tiny -} & {\tiny -}\\
{\tiny 28} & {\tiny \object{R\,Scl}} & {\tiny 2006-12-16 03:19:24} & {\tiny 0.13} & {\tiny G0-H0} & {\tiny 25} & {\tiny 93}  & {\tiny 1.4} & {\tiny SCI-PHOT} & {\tiny 30}\\ % P78 GTO
\hline
{\tiny Cal} & {\tiny \object{$\tau^{4}$\,Eri}} & {\tiny 2006-12-18 00:57:42} & {\tiny \ldots} & {\tiny -} & {\tiny \ldots} & {\tiny \ldots}  & {\tiny 1.2} & {\tiny -} & {\tiny -}\\
{\tiny 29} & {\tiny \object{R\,Scl}} & {\tiny 2006-12-18 01:21:29} & {\tiny 0.14} & {\tiny G0-H0} & {\tiny 31} & {\tiny 79}  & {\tiny 1.2} & {\tiny SCI-PHOT} & {\tiny 30}\\ % P78 GTO
\hline
{\tiny Cal} & {\tiny \object{$\tau^{4}$\,Eri}} & {\tiny 2006-12-21 01:27:43} & {\tiny \ldots} & {\tiny -} & {\tiny \ldots} & {\tiny \ldots}  & {\tiny 0.9} & {\tiny -} & {\tiny -}\\
{\tiny 30} & {\tiny \object{R\,Scl}} & {\tiny 2006-12-21 02:06:31} & {\tiny 0.15} & {\tiny G0-H0} & {\tiny 29} & {\tiny 86}  & {\tiny 1.7} & {\tiny SCI-PHOT} & {\tiny 30}\\ % P78 GTO
\hline
\end{tabular}
\end{table*}

The VINCI data (program 60.A-9220) were processed with the pipeline of data reduction based on the wavelet transform described in \citet{kervella04}. After checking the stability of the transfer function ($T_{i}$) over the two nights of observations, the raw visibility measurements were calibrated using the weighted average of the estimated $T^{2}_{i}$. \\
The VINCI instrument has a bandpass corresponding to the K-band filter (2.0-2.4 $\mu$m) without spectral dispersion. It is thus necessary to compute an effective wavelength to determine the spatial frequency of the observation \citep{kervella07}. The effective wavelength of our observations was calculated from the relation given by \citet{wittkowski04} using the best-fitting hydrostatic model spectrum of \object{R\,Scl} described in Sect.~\ref{stellar-atmosphere}. We obtained an effective wavelength of 2.253~$\mu$m.\\
Among the VINCI data, one observation (2001-12-08 01:03:54; see Table~\ref{journal-VINCI}) led to an outlier showing an abnormally low level of amplitude as compared to the other visibility data at close base, PA, and phase. The corresponding visibility measurement was discarded.\\
In the case of MIDI, chopped acquisition images were recorded (f=2\,Hz, 2000 frames, 4\,ms per frame) to ensure the accurate acquisition of the target. Photometry from the UT observations (program 074.D-0601) was obtained before and after the interferometric observations with the HIGH-SENS mode of MIDI, using the GRISM that provides a spectral resolution of about 230. Photometry from the AT observations were obtained with the HIGH-SENS mode (program 078.D-0112), and simultaneously with the interferometric observations with the SCI-PHOT mode of MIDI (programs 077.D-294 and 078.D-0122), using the PRISM that provides a spectral resolution of about 30. The acquisition images enabled us to confirm that the mid-IR source was unresolved by the single-dish UT. This implies that most of the mid-infrared flux originates in the inner 300 mas of the source, corresponding to the Airy pattern of the UTs and defining the Field Of View (FOV) of the interferometric observations. This also means that the flux detected within the 1.1$\arcsec$ FOV of the single-dish AT originates in this inner region. Therefore, measurements extracted from the ATs and the UTs are fully consistent with each other.\\
The data reduction software packages\footnote{\tt{http://www.mpia-hd.mpg.de/MIDISOFT/, http://www.strw.leidenuniv.nl/$\sim$nevec/MIDI/}} MIA and EWS \citep{jaffe04} were used to derive the calibrated spectra and visibilities \citep{chesneau05,ratzka07}. MIA is based on a power spectrum analysis and uses a fast Fourier transformation (FFT) to calculate the Fourier amplitude of the fringe packets, while EWS uses a shift-and-add algorithm in the complex plane, averaging appropriately modified individual exposures (dispersed channeled spectra) to obtain the complex visibility. \\ 
To calibrate the MIDI visibility measurements of \object{R\,Scl}, we used the transfer functions derived from the closest calibrators in time from the source (see Tables~\ref{journal-MIDI-UT} and \ref{journal-MIDI-AT}). This allows a decrease in the bias due to the air mass difference between the source observation at a given sky location and the calibrator observation at another location. The main source of uncertainties on the calibrated visibility measurements comes from the time-dependent atmospheric transmission between the observation of the calibrator and the source and from the dimension and uncertainty on the calibrator angular diameter \citep{borde02, cruzalebes10}. 
The rms of the transfer function deduced from all the calibrators of the night gives an upper estimate of the calibrated visibility error bar due the time-dependent atmospheric condition between the observations of the source and the calibrator.
We assume a standard error bar of 10\% on the calibrated visibilities corresponding to the upper limit on the error budget previously assigned. However, in the case of the HIGH-SENS mode, the photometry is performed about 5 to 10 minutes after the record of the fringe, i.e. correlated flux. As the atmospheric conditions change between both recordings, an additional error bar on the calibrated visibilities must be considered. Taking a similar error into account for both the SCI-PHOT and HIGH-SENS modes implicitly assumes that the variation in the transfer functions is \textit{much} greater than the intrinsic error of one measurement due to the changing weather conditions. During good atmospheric conditions, this assumption is reasonable. During bad atmospheric conditions (i.e. up to a certain seeing threshold), the large fluctuation of the positions of the two beams reduces the beam overlap leading to strong variations in the fringe amplitude and the photometry.
A solution to quantify the influence of the seeing on an individual measurement is to use the SCI-PHOT mode data with additional photometry (\textit{pseudo HIGH-SENS} mode) taken under similar atmospheric conditions. Comparing calibrated visibilities reduced with the pure SCI-PHOT mode with those reduced with the pseudo HIGH-SENS mode gives a direct handle on the midterm (i.e. time span from fringe tracking to the photometry) influence of the seeing. Under bad conditions (i.e. seeing$\ge$1.5\,$\arcsec$), the pseudo HIGH-SENS mode gives systematic lower visibilities, up to a level of 4\%, when comparing them to the ones deduced from the pure SCI-PHOT mode, while the difference falls to zero when the seeing is smaller than 1.1\,$\arcsec$. This is reasonable considering that the overlap of two beams is statistically reduced leading to systematically lower visibilities. We therefore consider an additional positive deviation of 4\% (i.e. up to the 5\% SCI-PHOT positive error) on the error bars associated to the calibrated visibilities recorded in HIGH-SENS mode under bad atmospheric conditions (i.e. seeing$\ge$1.5\,$\arcsec$). Finally, the error bar on the calibrated visibilities recorded in HIGH-SENS mode under better atmospheric conditions (i.e. seeing$\le$1.1\,$\arcsec$) is similar to the standard 10\% error bar assigned to the SCI-PHOT mode measurements.\\
The data quality has been checked using the \textit{MIA Graphical User Interface}, and the software developed by Paris Observatory for MIDI\footnote{Software is available through the JMMC website: \\ \tt{http://mariotti.ujf-grenoble.fr/}}. Among the MIDI measurements, data sets \#2, \#5, \#8, \#15, and \#25 (see Tables~\ref{journal-MIDI-UT} and \ref{journal-MIDI-AT}) led to outliers. Data sets \#2, \#15, and \#25 show calibrated visibilities greater than unity, data set \#5 shows an unusual spectral signature of the dispersed visibility, whereas data set \#8 shows an abnormally high level of visibility amplitude compared to the other visibility data at close base, PA, and phase. Finally data sets \#9, \#10, \#11, \#12, \#13, \#14, and \#16 show very broad dispersed power spectral densities (PSD) and/or fringe histograms (FH), especially from 8 to 10\,$\mu$m since shorter wavelengths are more affected by the atmospheric conditions. If the PSD and FH are very broad, a very significant fraction of the fringe power is distributed outside the integration range, leading to a systematic underestimation of visibility. All the corresponding visibility measurements were discarded.\\

Figure~\ref{VINCI-MIDI-uv_cov} shows the K-band $uv$ coverage of the VINCI observations (see Table~\ref{journal-VINCI}) and the N-band spectrally-dispersed $uv$ coverage of the MIDI observations (see Tables~\ref{journal-MIDI-UT} and \ref{journal-MIDI-AT}). The spatial coverage spreads over a large scale (from 15 to 60\,m baselines), whereas the angular coverage is limited to 90$^{\circ}$.\\

\begin{figure}[tbp]
\begin{center}
\includegraphics[width=9.0cm]{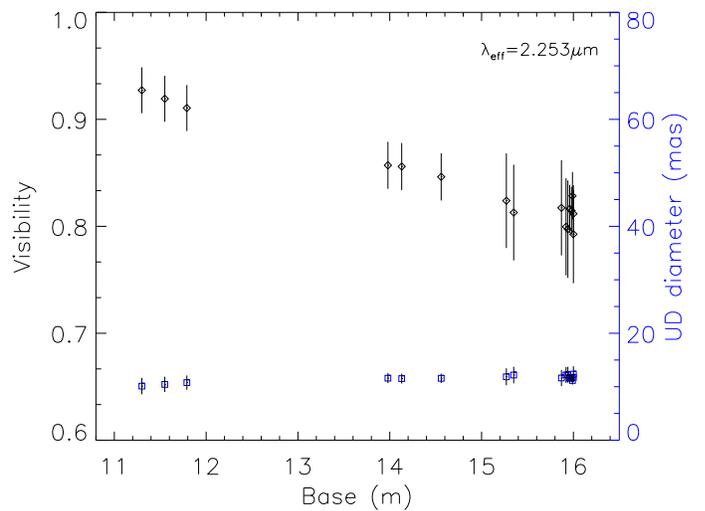}
\end{center}
\caption{Broadband VINCI visibilities of \object{R Scl} (black diamonds) as a function of the projected baselines observed over 2 nights between November and December 2001 (see Table~\ref{journal-VINCI}). Uniform disk diameters (blue squares to be read from the scale on the right axis) are computed from each visibility measurement.}
\label{VIS-VINCI-vs-PA}
\end{figure}

\begin{figure*}[tbp]
\begin{center}
\includegraphics[width=18.5cm]{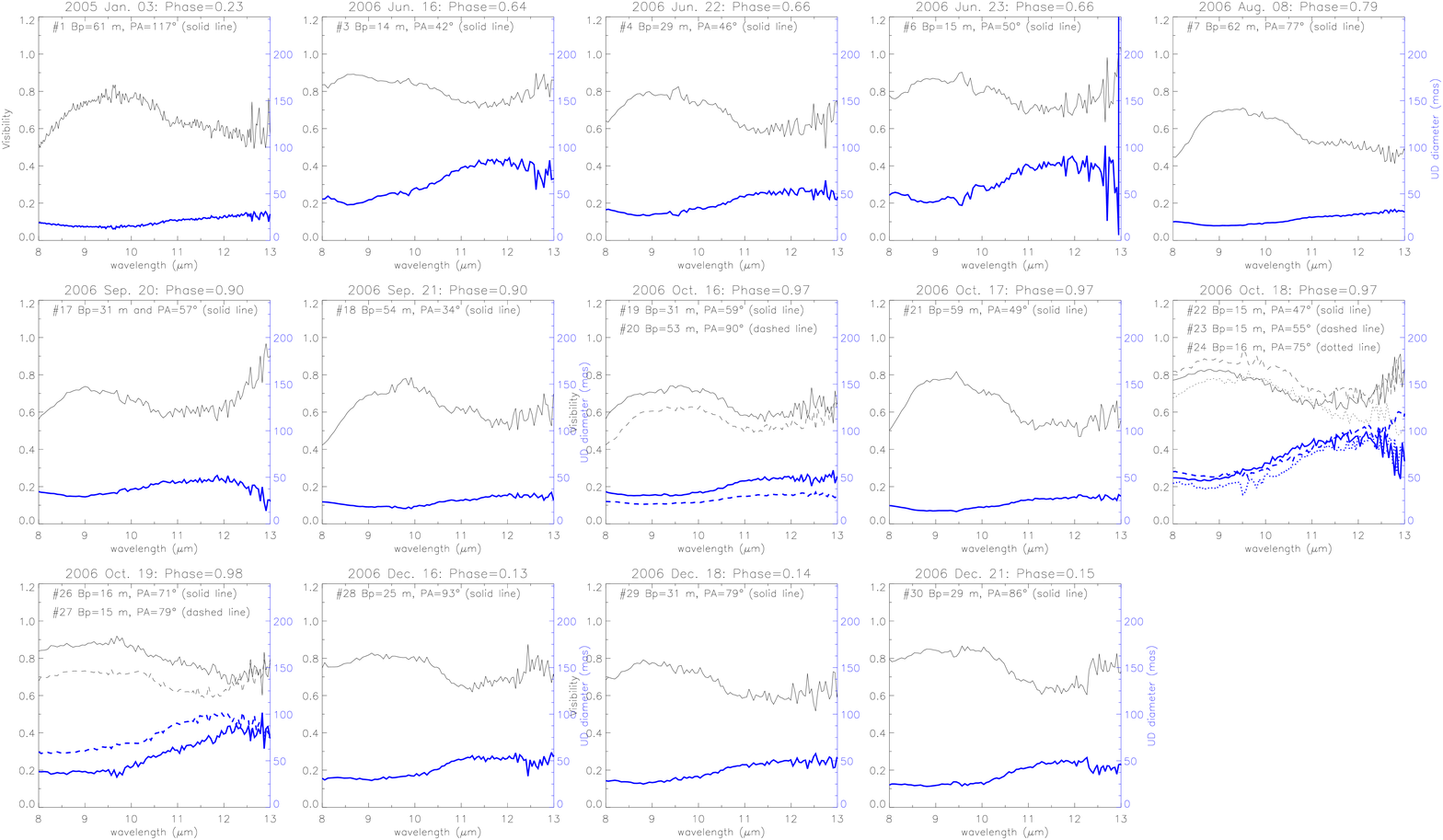}
\end{center}
\caption{Spectrally-dispersed MIDI visibilities of \object{R\,Scl} (black thin lines to be read from the scale on the left axis) observed over 14 nights between January 2005 and December 2006 (see Tables~\ref{journal-MIDI-UT} and \ref{journal-MIDI-AT}). Uniform disk diameters (blue thick lines to be read from the scale on the right axis) are computed from the visibilities at each spectral channel. Error bars are not included for clarity. The data set numbers are defined in Tables~\ref{journal-MIDI-UT} and \ref{journal-MIDI-AT}.}
\label{VIS-UD-MIDI-vs-lambda}
\end{figure*}

Figure~\ref{VIS-VINCI-vs-PA} presents the VINCI calibrated visibilities observed over 2 nights between November and December 2001 (see Table~\ref{journal-VINCI}). The angular diameters of the equivalent uniform disk are computed from each visibility measurement and show a stability from phase-to-phase at different position angles. Figure~\ref{VIS-UD-MIDI-vs-lambda} shows the MIDI calibrated visibilities observed over 14 nights between January 2005 and December 2006 (see Tables~\ref{journal-MIDI-UT} and \ref{journal-MIDI-AT}). The diameters of the equivalent uniform disk were computed from the visibilities at each spectral channel. The dimension of those diameters regularly increases from 9 to 13 $\mu$m, indicating an extended circumstellar environment \citep{ohnaka05}. \\

The phase shifts or differential phases correspond to the information extracted from the difference between the phase at a given wavelength and the mean phase determined in the full N-band region. It provides information about possible asymmetry of the source in the region covered by the different interferometric measurements. As an example, strong phase signals were detected in sources harboring dusty disks \citep{deroo07,ohnaka08}. Due to the low values ($<$\,$\pm$5$^\circ$) of the calibrated differential phases for each projected baseline, the global centrosymmetry of the close circumstellar environment of the object is confirmed in the mid-infrared for the corresponding angular coverage.

\section{Observed variability of \object{R\,Scl}}
\label{var}

Sections~\ref{photo-varia}, \ref{pola-varia}, and \ref{spectro-varia} discuss the variability of the star derived from photometry, polarimetry, and spectroscopy. Finally, Sect.~\ref{interfero-varia} is dedicated to the interferometric variability giving constraints on the dynamic process at work in the close circumstellar environment of \object{R\,Scl}.

\subsection{Photometric variability}
\label{photo-varia}

The K-band light curve of \object{R\,Scl} was discussed by \citet{whitelock97}. The authors distinguish two main periods of 374 and 1804 days. The shorter period is identified to be the pulsation period of the star, whereas the longer one might be a beat period between two similar pulsation periods \citep{houk63}. More recent polarimetric measurements (\citealt{yudin02}, see Sect.~\ref{pola-varia}) reveal a clumpy medium surrounding \object{R\,Scl} where orbiting dusty clump structures could also be the origin of the long period.\\ 

\begin{figure}[tbp]
\begin{center}
\includegraphics[width=9.0cm]{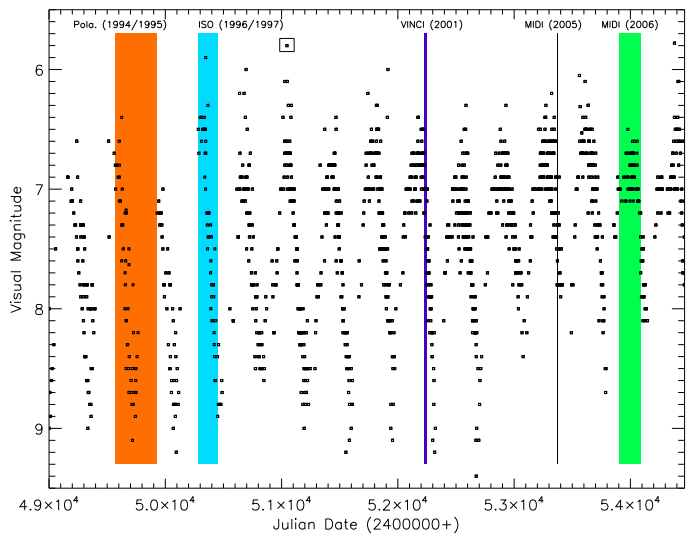}
\end{center}
\caption{AAVSO visual lightcurve of \object{R\,Scl} \citep{henden09}. Colored zones indicate the observing period of Polarimetric measurements (orange) \citep{yudin02}, ISO/SWS measurements (blue) \citep{sloan03}, VINCI measurements (violet), MIDI UT measurement (black), and MIDI AT measurements (green). The point surrounded by the open rectangle represents the chosen phase-zero point ($\phi_{0}$=0).}
\label{AAVSO_lightcurve}
\end{figure}

In the following, the phase of the star is established from the light curve according to

\begin{equation}
\label{eq-lightcurve}
\phi=\frac{(t - T_{0}) \, \rm mod \, P}{\rm{P}},
\end{equation}
where $t$ corresponds to the observing time in Julian day, $T_{0}$=2\,451\,044 is the Julian date of the selected phase-zero point corresponding to the maximum light of the star ($\phi_{0}$=0; see Fig.~\ref{AAVSO_lightcurve}), and P=374 days is the pulsation period of the star. \\

Besides the VINCI interferometric and MIDI spectro-interferometric observations of \object{R\,Scl}, polarimetric \citep{yudin02} and additional ISO/SWS spectrometric \citep{sloan03} data of the star are available.\\
Figure~\ref{AAVSO_lightcurve} represents the visual lightcurve of \object{R\,Scl} where the polarimetric (Pola.), spectroscopic (ISO and MIDI), and interferometric (VINCI and MIDI) observing periods are marked. 

\subsection{Polarimetric variability}
\label{pola-varia}

Wherever there is appreciable asymmetry in an astronomical situation, emitted light is likely to be polarized at some level. \citet{yudin02} show that \object{R\,Scl} displays polarimetric variability in the VR$_{C}$I$_{C}$ bands on different time scales from hours to one year. The authors argue that the polarimetric variations of the star are most probably not related to binarity, since these variations occur on timescales that are inferior to one pulsation period ($\sim$374 days). With a wavelength-dependence of polarization close to P$\propto$$\lambda^{4}$, consistent with Rayleigh scattering by small ($\approx$0.01 $\mu$m) amorphous grains, the authors suggest that the polarimetric variation would be related to dusty clumps located in the inner zone of the warm dusty shell (i.e. from 2 to 4~$R_{\star}$) with a quasi-period of rotation in the range 470 to 1600 days. \\  
The constraints provided by the MIDI differential phases (see Sect.~\ref{obs}), which do not exhibit any asymmetries in the close vicinity of the source (from 2 to 10~$R_{\star}$), contradict the conclusions of \citet{yudin02}. Therefore, we are left with three possibilities.\\

(i) \textbf{The MIDI interfererometric angular coverage is outside the dusty clumps region.}\\ 
It is difficult to determine the position of the clumps because of the large uncertainty on their rotational period found by \citet{yudin02}. As it is not possible to locate the clumps at the time of the MIDI observations, we cannot exclude that the clumpy region is situated out of the interferometric coverage, which only spreads over an angle of 90$^{\circ}$ (see Fig.~\ref{VINCI-MIDI-uv_cov}).\\ 

(ii) \textbf{The dynamic range of MIDI is not high enough to separate the emission of the dusty clumps from the one coming from the circumstellar material.}\\ 
We can investigate this point by an estimation of the constraints provided by the MIDI differential phase measurements. For that, we performed a perturbation of the intensity distribution of the best-fitting model (see Sect.~\ref{dust-chemical-composition}) by a Gaussian-like dust cloud of 1.8 mas width (defined from the density structure found by \citealt{yudin02}) varying its flux contrast, its distance (from 2 to 4~$R_{\star}$), and its position angle. Our tests allow any mid-IR dusty clump source representing more than about 4\% of the total flux of the object (i.e. 4 to 8~Jy between 8~$\mu$m and 13~$\mu$m) to be discarded. Taking the condensation density $\rho_{c}$ and the condensation mass derived by \citet{yudin02} into account, and describing the dusty clump as a blackbody of 1000 to 1500~K at a distance ranging from 2 to 4~$R_{\star}$, we obtain a flux range between 2.0 and 3.5~Jy in the N-band. As these values are below 4\% of the total flux of the object, this solution can be retained. \\

(iii) \textbf{The dusty clumps emission vanished from the N-band spectral window during the 10 years separation from the polarimetric to the MIDI interferometric observations.}\\
Shock waves could push the dusty clumps observed by polarimetry to regions where its emission peak is outside the mid-infrared window. However, taking a shell expansion velocity from 10.5 to 16.5 km\,s$^{-1}$ into account for \object{R\,Scl} \citep{schoier01,wong04}, the dusty clump would be located around 14.4 to 23.4~$R_{\star}$ (within the FOV of both AT and UT telescopes) at the end of the 10-year separation from the polarimetric to the MIDI observations. Assuming a Planckian radiation field geometrically diluted with distance from the star, the temperature of the clump reaches about 400~K corresponding to an emission peak well inside the mid-infrared window. Therefore this solution can be discarded.

\subsection{Spectrometric variability}
\label{spectro-varia}

\begin{figure}[tbp]
\begin{center}
\includegraphics[width=9.0cm]{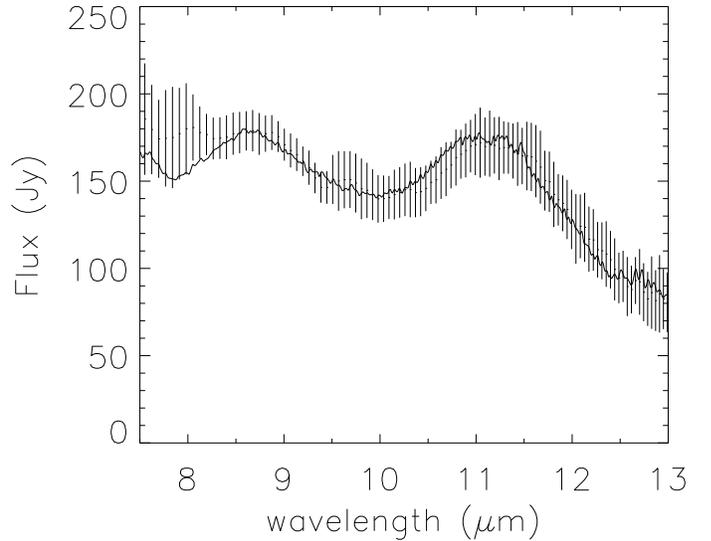}
\end{center}
\caption{Comparison of the ISO/SWS spectrum of \object{R\,Scl} at phase 0.97 (thick line), and the MIDI spectrum of \object{R\,Scl} at phase 0.98 (error bars).}
\label{Flux-MIDI-ISO-phase-000}
\end{figure}

\begin{figure}[tbp]
\begin{center}
\includegraphics[width=9.0cm]{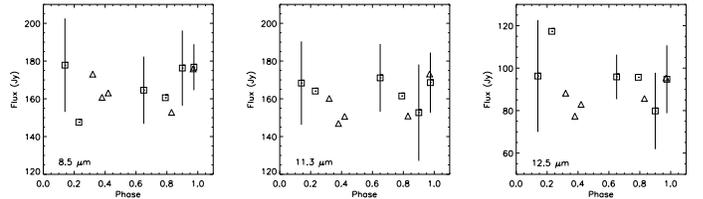}
\end{center}
\caption{MIDI (squares) and ISO/SWS (triangles) flux of \object{R\,Scl} as a function of the visual phase in the 8.5 (left), 11.3 (middle), and 12.5~$\mu$m (right) spectral band corresponding to the \textit{pseudo-continuum}, the SiC feature, and the C$_{2}$H$_{2}$+HCN feature, respectively.} 
\label{Flux-MIDI-ISO-all-phase} 
\end{figure}

As discussed in Sect.~\ref{obs}, most of the mid-infrared flux originates in the inner 300\,mas of the source. Therefore, the mid-infrared spectra obtained from the AT and UT single dishes are directly comparable to the ones produced by the ISO telescope (FOV=33$\arcsec$$\times$20$\arcsec$). To keep consistency in the comparison to the measurements, only ISO/SWS spectra using the AOT 1 template \citep{leech03} are considered in the following. Figure~\ref{Flux-MIDI-ISO-phase-000} compares both calibrated MIDI and ISO/SWS spectra \citep{sloan03} of \object{R\,Scl} around the maximum visual light ($\phi$$\sim$0).\\
Figure \ref{Flux-MIDI-ISO-all-phase} presents the calibrated MIDI and ISO/SWS flux of \object{R\,Scl} as a function of the phase at 8.5~$\mu$m (\textit{pseudo-continuum}: defined where the blackbody continuum fits to the ISO/SWS spectrum of the star), 11.3~$\mu$m (SiC feature), and 12.5~$\mu$m (C$_{2}$H$_{2}$+HCN) spectral bands \citep{hron98,gautschy04,zijlstra06}. Each monochromatic flux is determined by integration of the narrow-band filter fluxes over a bandwidth of $\Delta\lambda$=0.2~$\mu$m. The MIDI error bars are estimated from the standard deviation of the N-band flux-calibrated spectra determined from each night and each telescope, i.e. typically 3 to 10 spectra, except at phases 0.23 and 0.79 where only one spectrum is available (see Tables~\ref{journal-MIDI-UT} and \ref{journal-MIDI-AT}).\\

From Figs.~\ref{Flux-MIDI-ISO-phase-000} and \ref{Flux-MIDI-ISO-all-phase}, we can argue that the level of flux received by MIDI is equivalent to the one of ISO. It means that the cycle-dependent flux level of the object in the N-band has probably not changed significantly from the ISO/SWS (1996/1997) to the MIDI (2005/2006) observations. From Fig.~\ref{Flux-MIDI-ISO-all-phase}, it is rather difficult to determine a real trend in variability, keeping in mind typical errors in the absolute ISO/SWS flux level ranging from 10 to 12\% \citep{vanderbliek98} and an error of about 20\% for the MIDI flux.

\subsection{Interferometric variability}
\label{interfero-varia}

To study the interferometric variability of the star, we compare visibility measurements observed at the same projected baseline lengths for very close position angles (few degree differences) from one epoch to another one. The same projected baselines need to be used to probe the same spatial regions of the object, whereas very close projected angles are needed to avoid effect coming from potential deviation in the geometry of the object from spherical symmetry. \\

We first compare both sets of VINCI visibility measurements at $\phi$=0.17 and $\phi$=0.23 for the given 16\,m projected baseline length through a range of 5$^{\circ}$ in position angle. This comparison does not display any significant (larger than 2$\sigma$) temporal variation. This fact is, however, not surprising given the short time interval at post-maximum brightness of the star and because the broadband measurements will average out the possible short term effects of shock fronts in CO or CN lines (e.g. \citealt{nowotny05a}, \citealt{paladini09}). As the VINCI data only probe the very low spatial frequency, what remains are thus only the slow variations due to stellar pulsation (P=374 days).\\

\begin{figure}[tbp] 
\includegraphics[width=8.0cm]{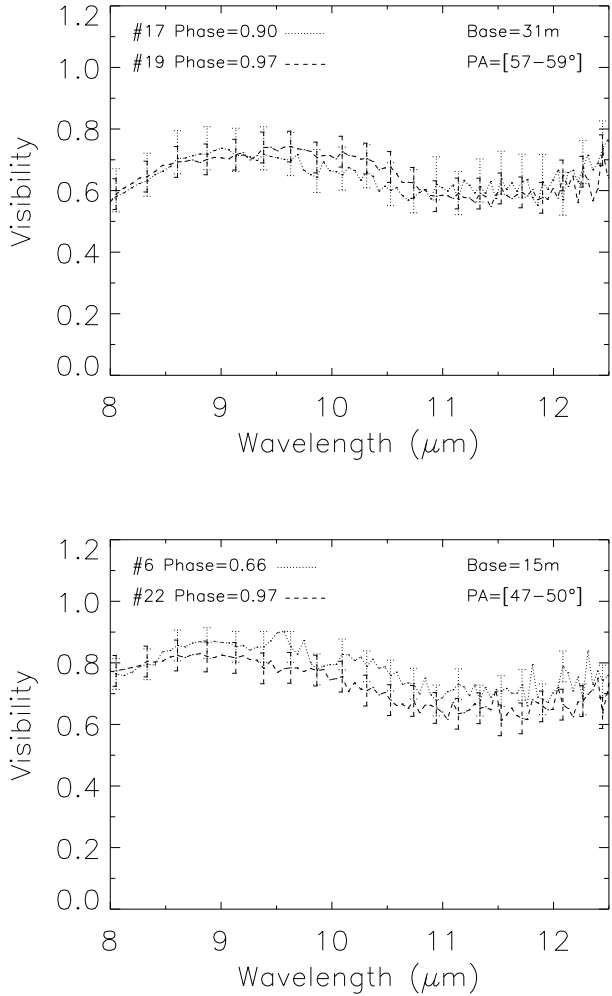} 
\caption{Comparison between spectrally-dispersed MIDI visibilities from phase-to-phase (black thin dashed and dotted lines) at a given projected baseline length for very close position angles. The data set numbers are defined in Table~\ref{journal-MIDI-AT}.}
\label{R-Scl-MIDI-interfero-varia}
\end{figure}

In the case of the MIDI measurements, the lowest spectral resolution (R=30) is high enough to resolve the broad features contained in the N-band spectral window (i.e. C$_{2}$H$_{2}$, HCN, C$_{3}$, and SiC). Therefore, we do not expect any effect that could possibly smear out the temporal variation coming from emission or absorption bands in that region. Figure~\ref{R-Scl-MIDI-interfero-varia} shows comparisons between the MIDI spectrally-dispersed visibilities from phase-to-phase at a given projected baseline length for very close position angles. The measurements do not show any significant variability effect for both 15 and 31\,m baselines. Because those baselines probe the region located from $\sim$17.5 to 100\,mas (3.4 to 19.6\,R$_{\star}$), this means that the stellar radiative pressure is not strong enough to reveal a significant movement of the warm mass dust shells detectable at the spatial resolution of the interferometer (from $\sim$8 to 12\,mas: 2.9 to 4.3\,AU at 360\,pc). Such behavior agrees with the amplitude of the mass shell of theoretical carbon-rich dynamic models, which does not exceed 3\,AU over a full cycle in the mid-infrared (see \citealt{paladini09}).

\section{Modeling the dynamic atmosphere of \object{R\,Scl}} 
\label{hydrodynamic_rscl}

The strategy of determining the best-fitting parameters for the dynamic model atmosphere of the C-rich star \object{R\,Scl} is the following:\\ 

\textbf{Step 1}: The hydrodynamic computation (see Sect.~\ref{vienna-model}) starts with an initial hydrostatic structure mainly characterized by the following parameters \citep{hoefner03}: luminosity, mass, effective temperature, metallicity, C/O ratio, and micro-turbulent velocity (for line opacities). In order to derive these parameters, we first perform a hydrostatic modeling of the stellar atmosphere (Sect.~\ref{stellar-atmosphere}) using a grid of COMARCS models \citep{aringer09}. By fitting synthetic spectra and visibilities based on such models to observed near-infrared spectro-interferometric measurements of \object{R\,Scl}, we find the best set of stellar atmospheric parameters for the initial hydrostatic structure of the dynamic model. \\

\textbf{Step 2}: The mid-infrared excess of \object{R\,Scl} shows an SiC feature around 11.2~$\mu$m (see Fig.~\ref{Flux-MIDI-ISO-phase-000}). This component must therefore be considered in the spectral synthesis for the comparison with observational data. Since only AmC dust is treated in the hydrodynamic calculation \citep{hoefner03}, opacities of SiC dust are added in the a posteriori radiative transfer computation (see Sect.~\ref{vienna-model}). Those opacities are determined by finding the proper SiC optical constants, and the fractional abundance of SiC to AmC dust that has condensed. These two characteristics are determined in Sect.~\ref{SiC-parameters} by comparing observed mid-infrared spectro-interferometric measurements of \object{R\,Scl} with synthetic spectra and visibilities derived from the DUSTY numerical code \citep{ivezic97,ivezic99}, which solves the radiative transfer equation through a dusty envelope.\\ 
In Sect.~\ref{disagree-model-data} we present the disagreement of the best-fitting model created by the combination of the hydrostatic COMARCS atmosphere and DUSTY circumstellar envelope (COMARCS+DUSTY model) with the observational data. This shows the necessity of using a self-consistent hydrodynamic model to understand how the time-dependent processes influence the atmospheric structures of the star. \\

\textbf{Step 3}: Finally, the self-consistent dynamic model used in Sect.~\ref{hydrodynamic-model-choice} is chosen by the parameters deduced from the best hydrostatic fit of the stellar atmosphere (see Sect.~\ref{stellar-atmosphere}) together with the SiC dust parameters derived from the DUSTY modeling (see Sect.~\ref{SiC-parameters}). We then interpret the spectro-interferometric data of \object{R\,Scl} using this model to discuss the dynamic picture deduced from the observations.

\subsection{Determination of the stellar parameters of \object{R\,Scl} with hydrostatic model atmospheres}
\label{stellar-atmosphere}

Besides the ISO/SWS spectra (see Sect.\ref{spectro-varia}), ground-based photometric measurements \citep{bagnulo98,whitelock06} are available for \object{R\,Scl} (see Table~\ref{photometry}). Using the relation of \citet{milne80}, also used by \citet{bagnulo98}, a visual interstellar extinction value of 0.18 was derived. The ratio of total-to-selective extinction at V, 3.1, on average, allowed us to correct the other magnitudes using the absorption curve of \citet{fitzpatrick99}. The data of \citet{whitelock06} are dispersed over a wide range of 26.3 pulsation periods of the star, whereas the data of \citet{bagnulo98} are measured at post-maximum light of the object.

\begin{table}[h]
\caption{Broadband photometry of \object{R\,Scl}} 
\label{photometry}
\begin{minipage}[h]{10cm}
\begin{tabular}
[c]{ccccc} \hline\hline
\tiny{Band} & \tiny{Extinction}  & \tiny{Dereddened} & \tiny{Phase} & \tiny{Reference} \\ 
            & \tiny{coefficient} & \tiny{magnitude} &  & \\ \hline
\tiny{B} & \tiny{0.25} & \tiny{11.03$\pm$0.3} \footnotemark[1]  & \tiny{0.17} & \tiny{\citet{bagnulo98}} \\
\tiny{V} & \tiny{0.18} & \tiny{4.55$\pm$0.7} \footnotemark[1]  & \tiny{0.17} & \tiny{-} \\ 
\tiny{J} & \tiny{0.05} & \tiny{1.56$\pm$0.05}  & \tiny{0.12} & \tiny{-} \\ 
\tiny{H} & \tiny{0.03} & \tiny{0.46$\pm$0.05}  & \tiny{0.12} & \tiny{-} \\ 
\tiny{K} & \tiny{0.02} & \tiny{-0.24$\pm$0.06}  & \tiny{0.12} & \tiny{-} \\
\tiny{L$^{\prime}$} & \tiny{0.01} & \tiny{-0.84$\pm$0.28}  & \tiny{0.12} & \tiny{-} \\\hline 
\tiny{J} & \tiny{0.05} & \tiny{1.97$\pm$0.86}  & \tiny{Whole cycle} & \tiny{\citet{whitelock06}} \\ 
\tiny{H} & \tiny{0.03} & \tiny{0.63$\pm$0.64}  & \tiny{-} & \tiny{-} \\ 
\tiny{K} & \tiny{0.02} &  \tiny{-0.10$\pm$0.36} & \tiny{-} & \tiny{-}\\ 
\tiny{L} & \tiny{0.01}  & \tiny{-0.74$\pm$0.28}  & \tiny{-} & \tiny{-} \\\hline
\end{tabular}
\end{minipage}
\footnotemark[1]{Uncertainties calculated from the measurements given in the \texttt{Simbad} database.}\\
\end{table}

\begin{figure}[tbp]
\begin{center}
\includegraphics[width=9.0cm]{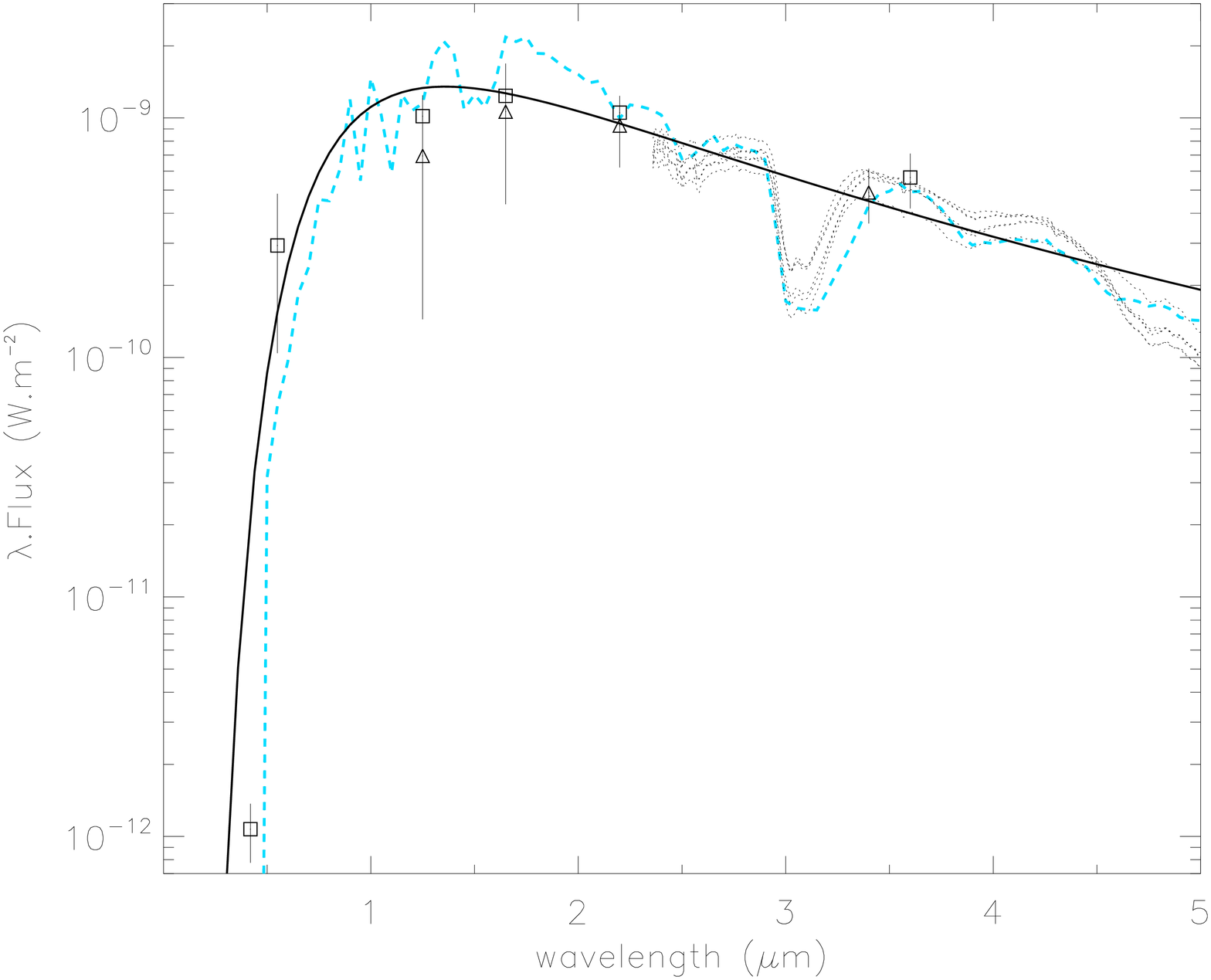} 
\end{center}
\caption{Best-fitting hydrostatic model (blue dashed line) on the spectrophotometric measurements. Photometric measurements (squares: \citealt{bagnulo98} + triangles: \citealt{whitelock06}; see Table~\ref{photometry}) and ISO/SWS spectra of \object{R\,Scl} at different epochs (dotted lines) are shown. A blackbody continuum having the same temperature (T$_{\rm eff}$=2700\,K) and angular diameter ($\oslash_{\rm \star}$=10.2\,mas) as the synthetic COMARCS spectrum is superimposed (black solid line).} 
\label{SED-MARCS-0--5mic}
\end{figure}

Because of all the absorption lines coming from strong molecular opacity sources (mainly species containing C, H, N, O; see \citealt{gautschy04}), spectra of cool stars differ a lot from a blackbody continuum spectra as is highlighted in Fig.~\ref{SED-MARCS-0--5mic}. The use of a more realistic synthetic carbon-rich spectrum is then required. The COMA code, developed by \citet{aringer00}, solves for the ionization and chemical equilibrium (using the method of \citealt{tsuji73}) at a given temperature and density (or pressure) combination for a set of atomic abundances assuming LTE. Based on these results, atomic and molecular opacities are calculated. Details of computation and information on recent updates on the COMA code can be found in \citet{lederer09}, and \citet{aringer09}. These opacities are then introduced in a spherically symmetric radiative transfer code to calculate the emergent intensity distribution. This intensity distribution is then used to derive the synthetic spectrum and visibility profile, which are compared with observational data of the star.\\ 

The determination of the stellar parameters is done by comparing hydrostatic COMARCS models to the spectrophotometric data up to 5~$\mu$m. This limit is fixed because of the contribution of the dust becoming non negligible at longer wavelengths for optically thin, \emph{warm dusty objects} like \object{R\,Scl} \citep{kraemer02}. This will be confirmed by the DUSTY modeling in Sect.~\ref{SiC-parameters}. The hydrostatic model was chosen as best-fitting of the broadband photometric data (see Table~\ref{photometry}) and the various features in the ISO/SWS spectra (see Fig.~\ref{SED-MARCS-0--5mic}) averaged over all the phases. Furthermore, the best-fitting model should also agree with the K-broadband VINCI interferometric measurements (see Sect.~\ref{obs}).\\
A least-square fitting minimization was completed from the grid of models included in \citet{aringer09}, varying the parameter values of effective temperature of the star from 2600 to 3000~K ($\Delta$T$_{\rm eff}$=100 K), the luminosity from 6000 to 8000~L$_{\odot}$ ($\Delta$L=200~L$_{\odot}$), the distance from 300 to 500\,pc ($\Delta$d=50\,pc), the mass for values of 1 and 2\,M$_{\odot}$, the metallicity for values of 1, 0.33, and 0.1~Z$_{\odot}$, and the C/O ratio for values of 1.1, 1.4, and 2.0. Finally, a value of 2.5~km\,s$^{-1}$ for the microturbulent velocity ($\zeta$), coming from the fit of very high spectral resolution data of carbon stars, is taken into account.\\
We obtained the best-fitting model to the ISO/SWS spectrometric data from 2.4 to 5~$\mu$m with a reduced $\chi^{2}$ of 3.5 (see Table~\ref{stellar-parameters}).\\ 

Figure~\ref{SED-MARCS-0--5mic} shows the best-fitting hydrostatic model on the spectrophotometric data of the star. As the VINCI data are integrated over the K-band (2.0-2.4~$\mu$m) defined by the corresponding filter curve, the synthetic broadband visibility results in a superposition of all the monochromatic profiles over the width of the band \citep{kervella03}. Figure~\ref{Visi-Kband-RScl-DUSTY} presents the best-fitting hydrostatic model superimposed on the VINCI visibility measurements.\\ 

Table~\ref{stellar-parameters} summarizes the stellar atmospheric parameters of \object{R\,Scl} deduced from the best-fitting hydrostatic model of the spectrophotometric and near-infrared interferometric measurements. \\

\begin{table}[h]
\caption{Stellar atmospheric parameters of \object{R\,Scl} deduced from best-fitting hydrostatic model of the spectrophotometric and near-infrared interferometric measurements.}
\label{stellar-parameters}
\centering
\begin{tabular}
[c]{cc} \hline\hline
Parameter & Value \\ \hline
Effective temperature: T$_{\rm eff}$ (K) & 2700 \\ 
Luminosity: L (L$_{\odot}$) & 7000 \\ 
Distance: D (pc) & 350 \\ 
Central star diameter: $\oslash_{\rm \star}$ (mas) & 10.2 \\
Surface gravity: log $g$ & -0.7 \\
Micro-turbulent velocity: $\zeta$ (km\,s$^{-1}$) & 2.5 \\
Stellar mass: M (M$_{\odot}$) & 1 \\
Metallicity: Z (Z$_{\odot}$) & 1 \\
Carbon over oxygen ratio: C/O & 1.4 \\ \hline
\end{tabular}
\end{table}

The values found agree well with the ones found in the literature. The distance of 350\,pc is relatively far from the Hipparcos distance of 474\,pc but in very good agreement with the distance of 360\,pc derived from the period-luminosity relation for carbon stars \citep{groenewegen96,whitelock08}. The angular diameter of the central star is relatively close to the one coming from the bolometric luminosity of \object{R\,Scl} (from 11.8 to 13.8 mas; \citealt{yudin02}). The C/O ratio of 1.34 derived by \citet{lambert86} agrees with the value that we found. The surface gravity, mass, metallicity, and C/O ratio are also in good agreement with those used by \citet{hron98} (log\,$g$=-0.5/-0.6, M=1\,M$_{\odot}$, Z=1\,Z$_{\odot}$, and C/O=1.4, respectively).

\subsection{Including of a dusty environment}
\label{dust-chemical-composition}

\subsubsection{Determination of the SiC dust parameters}
\label{SiC-parameters}

The DUSTY code \citep{ivezic97,ivezic99} solves the problem of radiation transport in a circumstellar dusty environment by calculating the radiative transfer equation in plane-parallel or spherical geometry. The radiation from the star is scattered, absorbed, and re-emitted by dust grains. DUSTY has built-in optical properties for the most common types of astronomical dust. The stellar spectrum can be specified in a numerical form as a separate user-supplied input file. In our case, the best-fitting COMARCS stellar atmosphere model, defined in Sect.~\ref{stellar-atmosphere}, is used as central source. The code can compute the wind structure by solving the hydrodynamic equations, including dust drift and the star's gravitational attraction, as a set coupled to radiative transfer \citep{ivezic99}. The calculation is performed for a typical wind in which the final expansion velocity exceeds 5~km~s$^{-1}$, in agreement with the expansion velocity found for \object{R\,Scl} from 10.5 to 16.5 km\,s$^{-1}$ \citep{schoier01,wong04}. This allows derivation of the mass-loss rate of the object, fixing the dust grain bulk density $\rho_{\rm s}$ and the gas-to-dust mass ratio r$_{\rm gd}$. \\
The input parameters of DUSTY related to the dusty circumstellar environment are the optical constants and the fractional abundances of the relevant dust grains, the grain size distribution, the overall radial optical depth at a given wavelength, the geometrical thickness, and the dust temperature at the inner boundary. Different types of grains can have different temperatures at the same location. However, DUSTY currently treats mixtures as single-type grains whose properties average the actual mix, therefore only one temperature is specified.\\

The determination of the parameters is done by fitting the COMARCS+DUSTY model to the photometric measurements, the ISO/SWS spectrometric data from 0.4 to 25~$\mu$m, and the MIDI spectrometric measurements from 8 to 13~$\mu$m, averaged over all the phases. Furthermore, the best-fitting model should also agree with the K-band VINCI and N-band MIDI interferometric measurements. For that, a least-square fitting minimization was completed first by testing different sets of optical constants for AmC and SiC grains and by computing a large grid of models varying the fractional abundances of the relevant dust grains ranging from 100$\%$ AmC to 100$\%$ SiC grains in 10$\%$ increments, the 0.55~$\mu$m overall radial optical depth from 0.05 to 1 ($\Delta$$\tau_{\rm 0.55 \mu m}$=0.05), and the dust temperature at the inner boundary from 800 to 1500~K ($\Delta$$T_{\rm in}$=100~K). As the geometrical thickness does not have any significant impact on the model spectrum because of the very optically thin circumstellar environment ($\tau_{\rm11.2\mu m}$=7.6$\times$10$^{-3}$, see Table~\ref{dusty-parameters}), we fixed this parameter to 1000 dust shell inner boundary radii knowing that the result is similar for values between 100 to 10$^{4}$. The grain size distribution was also fixed to a standard MRN grain size distribution as described by \citet{mathis77}\footnote{$n(a)$ $\propto$ $a^{-3.5}$ with minimum and maximum grain sizes of 0.005 $\mu$m and 0.25 $\mu$m, respectively}.\\
We obtained the best fit model to the ISO/SWS spectrometric data from 2.4 to 25~$\mu$m with a reduced $\chi^{2}$ of 3.5 (see Table~\ref{dusty-parameters}).\\

\begin{figure}[tbp]
\begin{center}
\includegraphics[width=9.0cm]{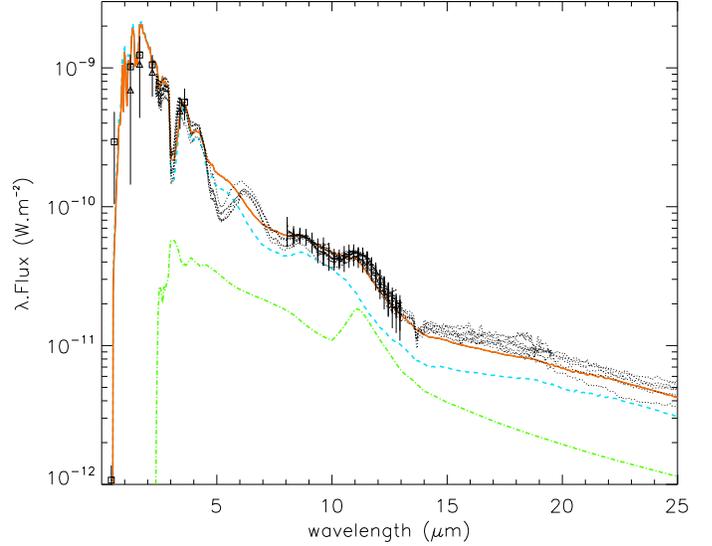}
\end{center}
\caption{Best-fitting COMARCS+DUSTY model (red solid line) on the ISO/SWS (dotted lines), MIDI (error bars) spectrometric measurements, and photometric (squares and triangles; see Table~\ref{photometry}) measurements of \object{R\,Scl}. The stellar contribution (blue dashed line) and the dust shell contribution (green dashed-dotted line) are also shown.}
\label{SED-MARCS-plus-DUSTY}
\end{figure}

\begin{figure*}[tbp]
\begin{center}
\includegraphics[width=18.0cm]{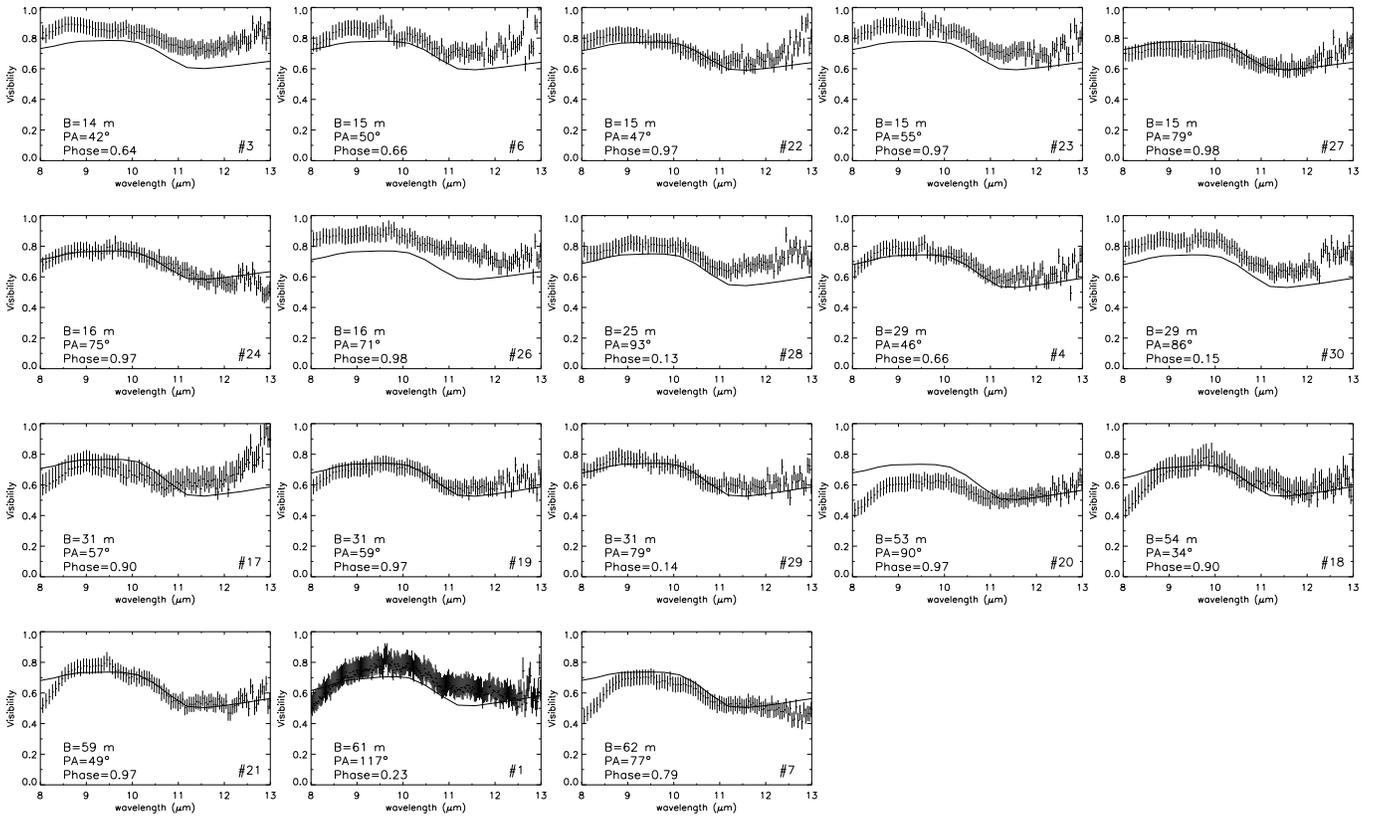}
\end{center}
\caption{Best-fitting COMARCS+DUSTY model (solid line) on the spectrally-dispersed MIDI visibilities of \object{R\,Scl} (error bars) from the smallest to the largest projected baselines. The data set numbers are defined in Tables~\ref{journal-MIDI-UT} and \ref{journal-MIDI-AT}.}
\label{Visi-lambda-RScl-DUSTY}
\end{figure*}

\begin{figure}[tbp]
\begin{center}
\includegraphics[width=9.0cm]{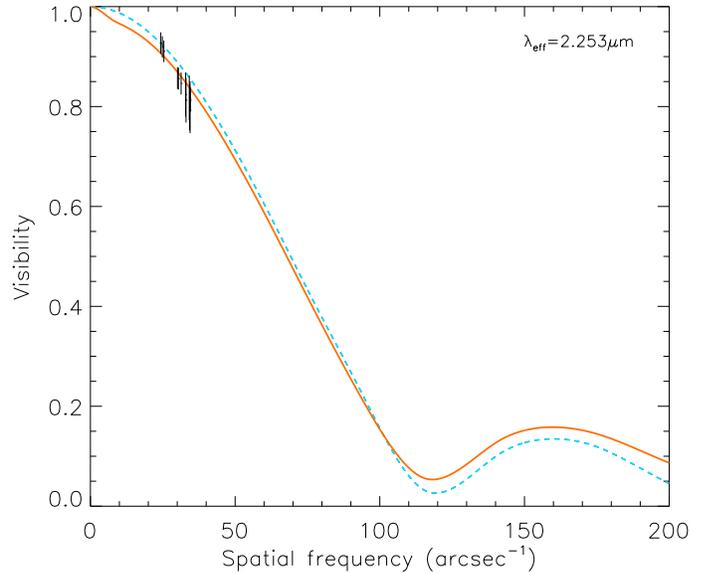}
\end{center}
\caption{Best-fitting hydrostatic model (blue dashed line) and best-fitting COMARCS+DUSTY model (red solid line) on the VINCI visibilities of \object{R\,Scl} (error bars).}
\label{Visi-Kband-RScl-DUSTY}
\end{figure}

Figure~\ref{SED-MARCS-plus-DUSTY} shows the best-fitting COMARCS+DUSTY model on the spectrophotometric data of the star. Figures~\ref{Visi-lambda-RScl-DUSTY} and \ref{Visi-Kband-RScl-DUSTY} present the best-fitting COMARCS+DUSTY model superimposed on the MIDI and VINCI visibility data, respectively. Table~\ref{dusty-parameters} summarizes the parameters of the dusty envelope found from the best-fitting COMARCS+DUSTY model on the spectrophotometric and the near- and mid-infrared interferometric measurements. These parameters are complementary to the stellar atmospheric parameters (see Table~\ref{stellar-parameters}).\\

\begin{table}[h]
\caption{Dusty circumstellar envelope parameters of \object{R\,Scl} deduced from the best-fitting COMARCS+DUSTY model on the spectrophotometric and the near- and mid-infrared interferometric measurements.}
\label{dusty-parameters}
\begin{minipage}[h]{10cm}
\begin{tabular}
[c]{cc} \hline\hline
Parameter & Value \\ \hline
Shell inner radius: $\varepsilon_{\rm in}$ (mas)  & 22.4 \\
Inner boundary temperature: T$_{\rm in}$ (K)  & 1200 \\
Grain chemical composition\footnotemark[1] & 90$\%$ AmC$^{r}$ + 10$\%$ SiC$^{p}$ \\
Grain size distribution\footnotemark[2] & MRN\\
Geometrical thickness: $Y$ (inner radii)& 1000 \\
Visual optical depth: $\tau_{\rm 0.55 \mu m}$ & 0.40 \\
1.0 $\mu$m optical depth: $\tau_{\rm 1.0 \mu m}$ & 0.18 \\
11.2 $\mu$m optical depth: $\tau_{\rm 11.2 \mu m}$ & 7.6$\times$10$^{-3}$ \\
Mass-loss rate\footnotemark[3]: \.{M} (M$_{\odot}$\,yr$^{-1}$) & (6.7$\pm$2)$\times$10$^{-7}$ \\\hline
\end{tabular}
\end{minipage}
\footnotemark[1]{AmC$^{r}$ stands for amorphous carbon of \citet{rouleau91} and SiC$^{p}$ for silicon carbide of \citet{pegourie88}.}\\
\footnotemark[2]{grain size distribution as described by \citet{mathis77}.}\\
\footnotemark[3]{assuming a dust grain bulk density $\rho_{\rm s}$=1.85\,g\,cm$^{-3}$ \citep{rouleau91} and a gas-to-dust mass ratio r$_{\rm gd}$=590 \citep{schoier05}.}
\end{table}

The resulting values agree well with the ones found in the literature. The inner radius of the dust shell is close to 5~$R_{\star}$ (see Table~\ref{stellar-parameters}) as determined by \citet{lorenz01}. The optical constants of AmC from \citet{rouleau91} and SiC from \citet{pegourie88}, together with the fractional abundance of SiC/AmC of 0.08 found by \citet{lorenz01}, are also in good agreement with the values found in this work. These last two characteristics allow us to determine the opacities of SiC used to interpret the spectro-interferometric data of \object{R\,Scl} with the help of the dynamic model (see Sect.~\ref{hydrodynamic-model-choice}). The AmC dust grain optical constants from \citet{rouleau91} are the same as adopted for the hydrodynamic calculation \citep{andersen03,hoefner03}. Good agreement is also found with the works of \citet{dehaes07} and \citet{lorenz01}, who determined a visual optical depth of 0.5$^{+0.2}_{-0.3}$ and a 1.0 $\mu$m optical depth of 0.1, respectively. The dust shell inner boundary temperature of 1000$^{+300}_{-200}$~K found by \citet{dehaes07} agrees well with the value determined in this work. Finally, the mass-loss rate of the star that we found also well agrees with the ones determined by \citet{lebertre97}: \.{M}=10$^{-7}$ M$_{\odot}$\,yr$^{-1}$; \citet{gustafsson97}: \.{M}=4$\times$10$^{-7}$ M$_{\odot}$\,yr$^{-1}$; and \citet{wong04}: \.{M}=2 to 5$\times$10$^{-7}$ M$_{\odot}$\,yr$^{-1}$. Such low values of mass-loss rate are consistent with the very optically thin circumstellar environment that we found from the fit ($\tau_{\rm 11.2 \mu m}$=7.6$\times$10$^{-3}$).

\subsubsection{Comparison of the COMARCS+DUSTY model with observations}
\label{disagree-model-data}

While the best-fitting COMARCS+DUSTY model is able to reproduce the largest part of the ISO/SWS spectra (see Fig.~\ref{SED-MARCS-plus-DUSTY}), we can see a discrepancy from 5 to 7~$\mu$m, which could be linked to wrong or incomplete C$_{3}$ line list opacities in that spectral range\footnote{Since there is only one C$_{3}$ line list available, a cross check of different line lists is not possible.}. Another discrepancy between the model and the ISO/SWS spectrometric data appears around the 13.7~$\mu$m C$_{2}$H$_{2}$ absorption feature. \citet{gautschy04} argue that this weak and sharp 13.7~$\mu$m absorption would come from cool layers above the 500~K dust and gas layer proposed by \citet{jorgensen00} to explain the lack of the 14~$\mu$m photospheric band.\\ 

The model matches the observed near- and mid-infrared visibilities data rather well (see Figs.~\ref{Visi-lambda-RScl-DUSTY} and \ref{Visi-Kband-RScl-DUSTY}), except for few measurements in the mid-infrared. The first disagreement comes from the amplitudes of the model visibilities, which are either higher or lower than the amplitudes of the visibility measurements at certain baselines. The most probable reasons for that follow.\\

(i) \textbf{The geometry of the object:}\\
Potential deviations from spherical symmetry (corresponding to the geometry of the COMARCS+DUSTY model) of the object at certain position angles could explain the disagreement. As the object subsequently appears more or less extended from one position angle to another, the corresponding amplitude of the visibility measurements decreases or increases, respectively, in contrast to the constant visibility amplitude of the COMARCS+DUSTY model.\\

(ii) \textbf{The variability of the object:}\\
It has been shown that the motion of the dynamic model atmospheres without mass loss mainly follows the pulsation of the stellar interior, whereas in the case of mass-losing models, the dust formation may deform the corresponding sinusoidal pattern (see Fig.~2 of \citealt{hoefner03}). As the object subsequently appears more or less extended from phase-to-phase, the corresponding amplitude of the visibility measurements decreases or increases \citep{paladini09}, respectively, in contrast to the constant visibility amplitude of the COMARCS+DUSTY model.\\

Another phenomenon that cannot be reproduced by the model is the slope in the N-band visibility data for the largest 60\,m projected baselines in the 8-9 $\mu$m spectral range (see Fig.~\ref{Visi-lambda-RScl-DUSTY}). As the model overestimates the visibility measurements, this reveals that the model is too optically thin. The regions probed at these baseline and spectral band are located between the photosphere of the star ($\sim$5\,mas; see Table~\ref{stellar-parameters}) and the dust shell inner radius ($\sim$22.4\,mas; see Table~\ref{dusty-parameters}). Because of the very low optical depth of the model in the 8-9~$\mu$m spectral range [$\bar{\tau}$(8-9~$\mu$m)=4.6$\times$10$^{-3}$], leading to domination of the stellar flux over the flux of the dust shell, the model is therefore close to the central star's intensity distribution. A way to reproduce the MIDI measurements is to increase the optical thickness of the model at the corresponding spatial frequencies. This can be done by adding extended molecular layers of C$_{2}$H$_{2}$ and HCN emitting in that spectral range and located above the stellar photosphere.\\ 
Many works in interferometry revealed that diameters of Mira and non-Mira M- and S-type stars appear systematically larger than expected in the near-infrared \citep{mennesson02,ohnaka04,perrin04,weiner04,verhoelst06} and in the mid-infrared \citep{weiner00,weiner03a,weiner03b,weiner04,ohnaka04,ohnaka05,verhoelst06,sacuto08}. This increase cannot be attributed only to dust shell features, but also to the possible existence of extended gaseous shells above the stellar photosphere. Such a molecular shell, also called \textit{MOLsphere} \citep{tsuji00}, is favored by the low surface gravity (log $g$=-0.7) and the high luminosity ($\sim$7000 $L_{\odot}$) of the star, allowing the levitation of the upper layers by shock waves or other dynamic phenomena in an easiest way. The temperatures in this levitated matter are low enough that large amounts of polyatomic molecules can form \citep{hron98,woitke99,gautschy04,verhoelst06}. \citet{ohnaka07} present their results on the modeling of the MIDI measurements of the carbon-rich Mira variable \object{V\,Oph}. The model consists of optically thick warm and cool C$_{2}$H$_{2}$ molecular layers and an optically thin AmC+SiC dust shell.\\

Although this \textit{MOLsphere} is useful for increasing the optical thickness of the close circumstellar structure, it remains an inconsistent ad-hoc model. In the classical \textit{MOLsphere} scenario, as is usually applied to interpreting interferometric or spectroscopic data, it is assumed that there are one or a few static layers with a certain concentration and temperature of a molecular species. The possible physical status and origin of such a layer is left unanswered, which is problematic for the following points.\\
(i) The radial temperature-density distribution of gas and dust around evolved objects is always a continuous and inconstant phenomenon.\\
(ii) The molecular (and dust) concentrations in a layer depend on each other, so it is impossible to freely choose the amount of certain species.\\
(iii) The radial structures are extremely time dependent.\\

As shown by \citet{hron98}, \citet{woitke99}, and \citet{gautschy04}, the hydrodynamic modeling leads to a self-consistent formation of extended molecular structures located between the photosphere and the dust formation zone. Finally, \citet{hoefner03} argue that sequences of hydrostatic models with varying stellar parameters cannot reproduce the effects of shock waves, levitation, or winds. The only way to understand how the dynamic processes influence the atmospheric structure on different spatial scales is the use of time-dependent dynamic atmospheric modeling.

\subsection{Dynamic model atmosphere for \object{R\,Scl}} 
\label{hydrodynamic}

The aim of this section is to show that a self-consistent hydrodynamic modeling forming a gaseous and dusty environment by itself is able to reproduce the time-dependent signature of the spectro-interferometric data of a carbon-rich, long-period variable star. Section~\ref{vienna-model} presents the dynamic model atmosphere used for the interpretation. Section \ref{bolo-visual-phase} connects the bolometric phases given by the model with the visual phases of the observations which is needed to compare time-dependent models with real measurements of \object{R\,Scl}. Section \ref{hydrodynamic-model-choice} presents the specific dynamic atmospheric model that we chose for the comparison with the spectro-interferometric data of the star. Section \ref{hydrodyn-molsphere} shows the ability of the dynamic model to form molecular layers located above the photosphere of the star in a self-consistent way. Finally, the last section (Sect.~\ref{hydrodynamic-comparison}) compares the dynamic model with the time-dependent spectro-interferometric data.

\subsubsection{Presentation of the hydrodynamic model atmospheres}
\label{vienna-model}

The self-consistent hydrodynamic structures for dust-forming atmospheres and winds used here are taken from the models contained in \citet{mattsson08} and from additional models required for the present work (Mattsson, priv. comm.). The dynamic model atmosphere is based on the code described by \citet{hoefner03} and \citet{mattsson10}. The models were obtained by solving the coupled system of hydrodynamics and frequency-dependent radiative transfer in spherical symmetry, together with a set of equations describing the time-dependent formation and the growth and evaporation of dust grains. The formation of AmC dust is treated in a self-consistent way using the moment method described by \citet{gail88} and \citet{gauger90}. Optical properties of AmC grains were taken from \citet{rouleau91}. Each dynamic model starts from a hydrostatic initial structure, and the effects of stellar pulsation are simulated by a variable inner boundary just below the stellar photosphere (piston model). The dynamic code provides temporally varying temperature and density stratifications of the object. Based on such radial structures, at given instances of time, opacities are computed using the COMA code (see Sect.~\ref{stellar-atmosphere}). The resulting opacities are used as input for a spherical radiative transfer code that computes the emergent intensity distribution for every frequency point of the calculation. This is used to calculate the synthetic spectra and visibility profiles, which are compared to the observational data of \object{R\,Scl}. Since only AmC dust is treated in the hydrodynamic calculation, opacities of SiC dust are added in the a posteriori radiative transfer computation. Those opacities are determined with COMA taking SiC optical constants into account and assuming that SiC dust is scaled with AmC dust. The scaling factor corresponds to the fractional abundance of SiC to AmC dust, which has condensed. Finally, SiC dust is assumed to follow the same temperature-density structure as AmC dust, derived from the full dynamic calculation.

\subsubsection{Bolometric and visual phases}
\label{bolo-visual-phase}

The phases of the dynamic model correspond to the bolometric phases derived from its bolometric lightcurve \citep{nowotny05b}. The phase-zero point was chosen at the maximum bolometric luminosity of the given dynamic model. As the radiation-hydrodynamic code stores atmospheric structures at more or less randomly distributed instances of time, the uncertainty in the choice of the phase-zero point amounts to a range of $\pm$0.03.\\
For O-rich Miras, a small shift between bolometric and visual lightcurves probably exists, such that bolometric phases lag behind visual ones by a value of $\approx$\,0.1 \citep{lockwood71,nowotny10}. This phase shift is caused by the behavior of molecules occurring in atmospheres of red giants and their corresponding spectral features in the visual. Recently, \citet{nowotny10} have studied the photometric variation in the $V$-filter for similar dynamic model atmospheres as the ones used here. The authors find that, for models developing no wind such as their model W or the one used in this paper to represent \object{R\,Scl} (see Sect.~\ref{hydrodynamic-model-choice}), the visual lightcurves follow the bolometric light variation caused by the variable inner boundary (piston) closely. Therefore, in the following, bolometric phases of the model within a range of $\pm$0.03 around the observed visual phases of \object{R\,Scl} are used for comparisons between model and observations.

\subsubsection{Choice of a specific dynamic model}
\label{hydrodynamic-model-choice}

The choice of the best dynamic model for \object{R\,Scl} comes from the stellar parameters and the distance derived from the previous hydrostatic modeling (see Sect.~\ref{stellar-atmosphere}). \\

Among the dynamic models that we have, one model shows parameter values of the hydrostatic initial structure close to the ones we derived from the hydrostatic COMARCS modeling (see Table~\ref{stellar-parameters}). Table~\ref{hydrodynamic-models-table} compares the stellar parameters of the hydrostatic model initiating the hydrodynamic computation with the stellar parameters derived from the best-fitting hydrostatic model determined in Sect.~\ref{stellar-atmosphere}. \\  

Opacities of SiC dust are added in the a posteriori radiative transfer calculation using the SiC optical constants of \citet{pegourie88} and a fractional abundance of SiC to AmC dust of 10\%, as derived from the COMARCS+DUSTY modeling (see Sect.~\ref{SiC-parameters}).\\

\begin{table}[h]
\centering
\caption{Parameter values of the hydrostatic initial structure of the hydrodynamic model used in this work (first row) compared to the parameters derived from the best-fitting hydrostatic model determined in Sect.~\ref{stellar-atmosphere} (second row).}
\label{hydrodynamic-models-table}
\begin{tabular}
[c]{ccccccc} \hline
Model & L$_{\rm \star}$ & M$_{\rm \star}$ & T$_{\rm \star}$ & Z$_{\rm \star}$ & $\zeta$ &  C/O  \\  
      & (L$_{\odot}$)   & (M$_{\odot}$)    &  (K)          & (Z$_{\odot}$)   & (km\,s$^{-1}$)  \\ \hline
Hydrodynamic & 7080 & 1 & 2800 & 1 & 2.5 & 1.35 \\ \hline 
Hydrostatic & 7000 & 1 & 2700 & 1 & 2.5 & 1.40 \\ \hline 
\end{tabular}
\end{table}

Two other free parameters are required for the hydrodynamic computation (see Sect.~\ref{vienna-model}):\\

\textbf{P$_{\rm mod}$: the pulsation period of the piston.}\\
A pulsation period of P$_{\rm mod}$=390~days, the closest to the pulsation period of \object{R\,Scl} (374~days), was used.\\

\begin{figure}[tbp]
\begin{center}
\includegraphics[width=9.0cm]{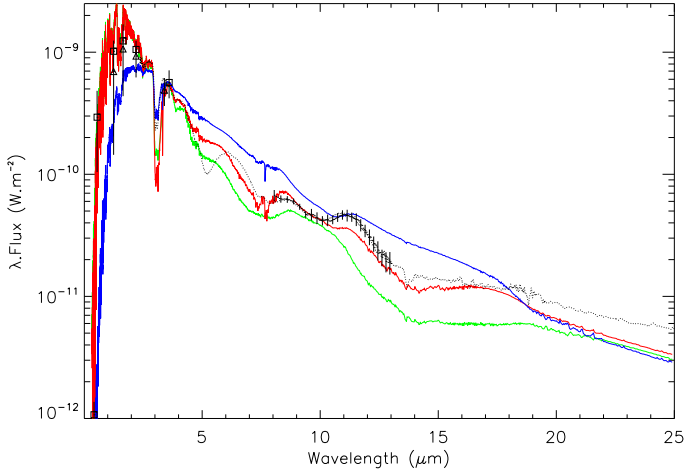}
\end{center}
\caption{Synthetic spectra based on the dynamic models at phase 0.98 with piston velocities of 2 (green line), 4 (red line), and 6~km~s$^{-1}$ (blue line), keeping the same parameter values as the ones contained in Table~\ref{hydrodynamic-models-table}. The dotted line is the ISO/SWS spectrum of the star at phase 0.97, whereas the error bars in the N-band correspond to the MIDI fluxes at phase 0.98. Squares and triangles correspond to the photometric measurements of the star (see Table~\ref{photometry}).}
\label{Flux-hydrodyn-diff-piston}
\end{figure}

\textbf{$\Delta$u$_{\rm p}$: the amplitude of the piston velocity.}\\
Different amplitudes of the piston velocity were tested from a range of $\Delta$u$_{\rm p}$=2 to 6~km\,s$^{-1}$, with steps of 1~km\,s$^{-1}$ (except for the $\Delta$u$_{\rm p}$=3~km\,s$^{-1}$ model not at our disposal).\\
Determination of the best-fitting piston velocity amplitude is done by comparing the synthetic spectra based on these models to the spectrophotometric data. Increasing the amplitude of the piston moves the material to larger distances from the star, which favors dust formation and reduces the effective gravitation (e.g. \citealt{winters00}).\\
For a piston velocity lower than 4~km s$^{-1}$ (i.e. model with $\Delta$u$_{\rm p}$=2~km\,s$^{-1}$ in Fig.~\ref{Flux-hydrodyn-diff-piston}), it appears that the model shows too low a level of flux compared to the spectrometric data. Furthermore, as already discussed in \citet{gautschy04}, a broad and intense C$_{2}$H$_{2}$/HCN absorption band around 14~$\mu$m is present, in pronounced contrast to the observed ISO/SWS spectra which show a continuum-like infrared excess with a weak and narrow C$_{2}$H$_{2}$ feature at 13.7~$\mu$m (see Fig.~\ref{Flux-hydrodyn-diff-piston}). Finally, the variability amplitude of the model is smaller than 0.3 in the V magnitude, as compared with the 1.4 variability amplitude of the cycle-to-cycle averaged visual magnitudes of the star. Therefore, in the case of too low a piston velocity, the dynamic model corresponds to a dust-free pulsating atmosphere without any wind (no mass-loss), leading to a pure photospheric spectrum such as the one shown by the blue dashed line of Fig.~\ref{SED-MARCS-plus-DUSTY}. \\   
When the piston velocity amplitude exceeds 4~km s$^{-1}$ (i.e. model with $\Delta$u$_{\rm p}$=6~km\,s$^{-1}$ in Fig.~\ref{Flux-hydrodyn-diff-piston}), the dynamic model shows too high a level of flux compared to the spectrometric data and a filling up of the whole molecular features at wavelengths longwards of 3.5~$\mu$m. Furthermore, the variability amplitude of the model is greater than 5.5 in the V magnitude, as compared with the 1.4 variability amplitude of the cycle-to-cycle averaged visual magnitudes of the star. In that case, the dynamic model develops a strong wind (\.{M}$>$10$^{-6}$ M$_{\odot}$\,yr$^{-1}$) where the large amount of formed gas and dust, located in the outer layers, obscures the innermost structures of the stellar atmosphere, although not significantly enough to fill up the 3.05~$\mu$m feature. \\ 
A dynamic model with a piston velocity of 4~km s$^{-1}$ gives the best overall match of the spectrophotometric data from 0.4 to 25~$\mu$m, even if the emission is still lower than the [13-25~$\mu$m] infrared excess of the ISO/SWS spectrum (see Fig.~\ref{Flux-hydrodyn-diff-piston}). The variability amplitude of the model of about 1 in the V magnitude is relatively close to the 1.4 variability amplitude of the cycle-to-cycle averaged visual magnitudes of the star. In that case, the model is, in some sense, a transition case between a simple, periodically pulsating dust-free atmosphere and an object featuring a dusty wind. The kinetic energy supplied by the pulsation (piston) causes strong, quasi-periodic atmospheric dynamics and some intermittent dust formation, but is not sufficient to start a dust-driven outflow.\\
The possible reasons for the discrepancy in the 5 to 7~$\mu$m wavelengths regions are similar to the ones expressed in Sect.~\ref{disagree-model-data} suggesting wrong or incomplete C$_{3}$ line list opacities in that spectral range. The model is also unable to reproduce the weak observed 3.05~$\mu$m C$_{2}$H$_{2}$/HCN feature.\\
Possible explanations of the discrepancy between the model and the measurements are discussed in more detail in Sect.~\ref{hydrodynamic-comparison}. This last model is, however, considered as the best dynamic model for \object{R\,Scl}, among the ones available, and will be used in the comparison to the observed spectro-interferometric measurements of the star in the following.\\

\subsubsection{Self-consistent formation of extended molecular structures}
\label{hydrodyn-molsphere}

\begin{figure}[tbp]
\begin{center}
\includegraphics[width=9.0cm]{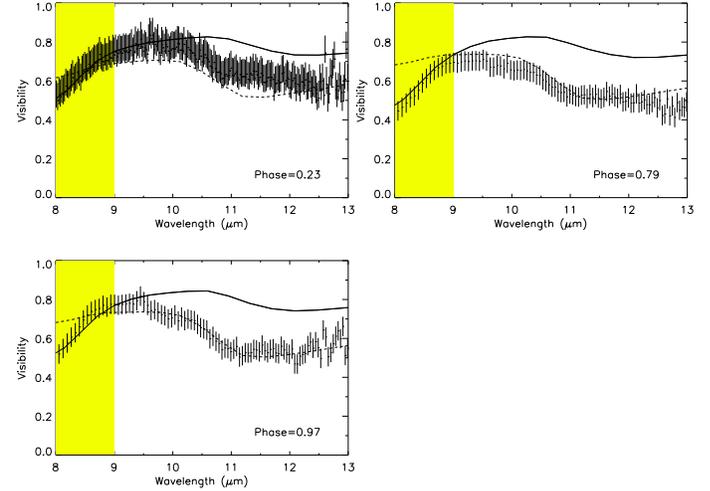}
\end{center}
\caption{Comparison of the visibility of the best dynamic model (solid line) with the best-fitting COMARCS+DUSTY model (dashed line; see Sect.~\ref{SiC-parameters}) superimposed on the 60\,m MIDI visibility data (error bars) at phases 0.23/0.79/0.97 (data sets \#1, \#7, and \#21, respectively). The yellow zone corresponds to the region dominated by warm molecular layers.} 
\label{VIS-MOLsphere-hydrodyn}
\end{figure}

As discussed in Sect.~\ref{disagree-model-data}, the N-band visibility data display a slope for the largest 60\,m baseline in the 8-9~$\mu$m spectral range likely related to the emission of extended molecular structures of C$_{2}$H$_{2}$ and HCN. Figure~\ref{VIS-MOLsphere-hydrodyn} shows the comparison of the visibility of the best dynamic model for a given cycle with the best-fitting COMARCS+DUSTY model superimposed on the MIDI visibility data around 60\,m for all the available phases: 0.23/079/0.97. From this figure, we can clearly see that the COMARCS+DUSTY model is too optically thin, showing a flat visibility profile in the 8-9~$\mu$m spectral range (see Sect.~\ref{disagree-model-data}). On the contrary, the dynamic model is able to reproduce the slope of the MIDI visibilities well through this wavelength range. This means that the extension of the dynamic model is suitable for revealing the presence of the molecular structures of C$_{2}$H$_{2}$ and HCN, emitting in that spectral range and located above the stellar photosphere. However, the dynamic model is not optically thick enough to reproduce the measurements in the reddest part of the mid-infrared spectral band.

\subsubsection{Comparison of the dynamic modeling results to the time-dependent spectro-interferometric data}
\label{hydrodynamic-comparison}

In the following, we investigate 100 phases of the model randomly distributed over 56 cycles corresponding to a time sequence of 60 years. As there is no possibility to define the model cycle corresponding to the observations, we decide to follow a statistical approach and to represent all the available cycles for a phase range close to the observed one (see Sect.~\ref{bolo-visual-phase}).\\

\begin{figure*}[tbp]
\begin{center}
\includegraphics[width=18.0cm]{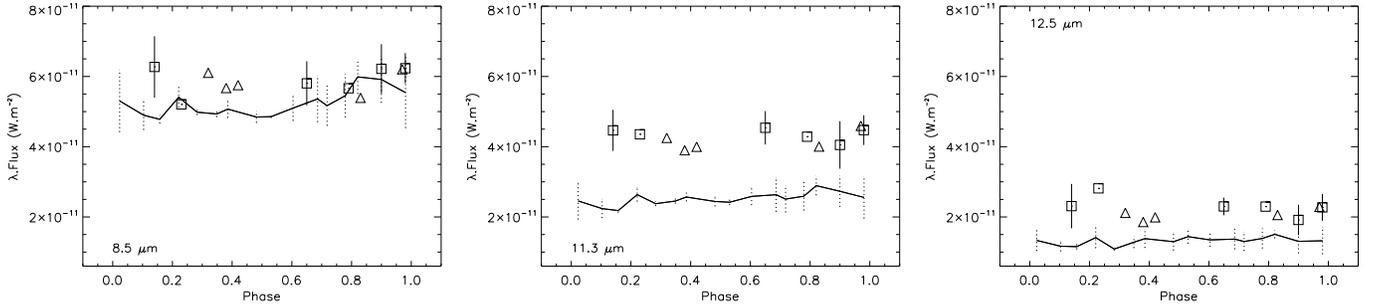}
\end{center}
\caption{Time-dependent cycle-to-cycle averaged fluxes of the best dynamic model (solid line) superimposed on the MIDI (squares) and ISO/SWS (triangles) spectrometric data in the 8.5 (left), 11.3 (middle), and 12.5~$\mu$m (right) spectral bands. Dotted error bars give the dispersion created by the cycle-to-cycle variation.}
\label{Flux-Hydrodyn-vs-Phase}
\end{figure*}

Figure~\ref{Flux-Hydrodyn-vs-Phase} shows the time-dependent cycle-to-cycle averaged fluxes of the best dynamic model superimposed on the MIDI and ISO/SWS spectrometric data in the 8.5, 11.3, and 12.5~$\mu$m spectral bands. The figure shows that the model agrees relatively well with the time-dependent flux data at 8.5~$\mu$m, whereas the model is fainter than the measurements in the 11.3 and 12.5~$\mu$m spectral bands.\\

In compliance with the observations, the comparison of the cycle-to-cycle averaged visibilities of the best dynamic model from phase-to-phase, considering the same baselines and phases as the observations of \object{R\,Scl} (see Sect.~\ref{interfero-varia}), does not show any significant variability of the object in the K and N bands.\\

\begin{figure}[tbp]
\begin{center}
\includegraphics[width=9.0cm]{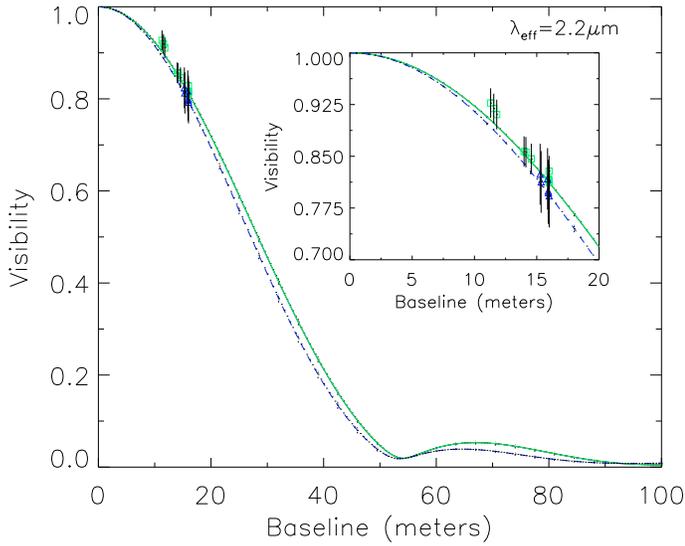}
\end{center}
\caption{Cycle-to-cycle averaged visibility profiles of the best dynamic model (green solid line: phase 0.17; blue dashed line: phase 0.23) superimposed on the VINCI data at phase 0.17 (green squares) and phase 0.23 (blue triangles). Dotted error bars give the dispersion created by the cycle-to-cycle variation. In the inset, a zoom of the relevant frequency region is represented.}
\label{VIS_Hydrodyn_VINCI}
\end{figure}

Figure~\ref{VIS_Hydrodyn_VINCI} shows the cycle-to-cycle averaged visibility of the best dynamic model superimposed on the VINCI data at phases 0.17 and 0.23. The model shows good agreement with the data implying that the extension of the model is suitable in the K-band. \\ 

\begin{figure*}[tbp]
\begin{center}
\includegraphics[width=18.0cm]{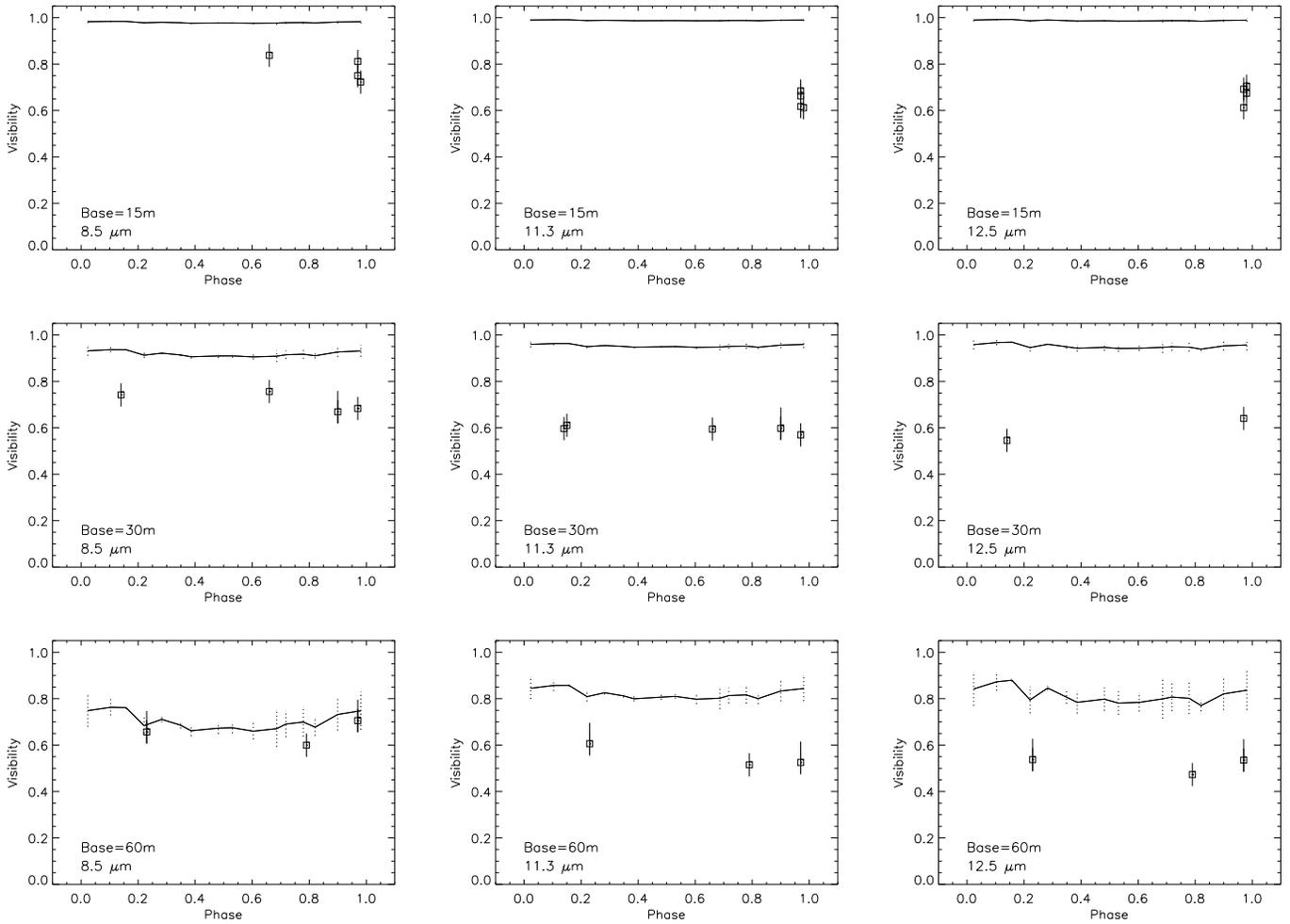} 
\end{center}
\caption{Time-dependent cycle-to-cycle averaged visibilities of the best dynamic model (solid line) superimposed on the [14-16\,m], [29-31\,m], and [59-62\,m] baselines MIDI data (squares) in the 8.5 (left), 11.3 (middle), and 12.5~$\mu$m (right) spectral bands. Dotted error bars give the dispersion created by the cycle-to-cycle variation.}
\label{VIS_Hydrodyn_MIDI}
\end{figure*}

Figure~\ref{VIS_Hydrodyn_MIDI} shows the time-dependent cycle-to-cycle averaged visibilities of the best dynamic model superimposed on the [14-16\,m], [29-31\,m], and [59-62\,m] baselines MIDI data in the 8.5, 11.3, and 12.5~$\mu$m spectral bands. The figure only presents measurements showing a deviation from the best-fitting COMARCS+DUSTY model (see Fig.~\ref{Visi-lambda-RScl-DUSTY}) smaller than the object variability scattering at 15 and 31\,m baselines (see Fig.~\ref{R-Scl-MIDI-interfero-varia}). This allows us to avoid comparing the spherically symmetric model with visibility measurements exhibiting a potential departure of the object from sphericity. As we do not have any information on the variability effect at the 60\,m baseline, all the measurements are represented. From this figure, it is obvious that the model is not extended enough to reproduce the level of the mid-infrared interferometric data. Only the visibility measurements at the highest spatial frequencies (B=60m and $\lambda$=8.0-9.5~$\mu$m), probing regions corresponding to the warm extended molecular layers (from 1.5 to 2~R$_{\star}$), are in good agreement with the model.\\

Besides the possibility of an inaccurate estimation of the parameter values of the hydrostatic initial structure, the most probable reasons for the discrepancy between the dynamic model and the spectro-interferometric measurements are the following ones.

\bigskip

(i) \textbf{The sparse sampling of the transition region from windless models to models with considerable outflows in certain critical stellar parameters:}\\
The 100~K higher value of the dynamic model effective temperature compared to the hydrostatic model could be a reason for the discrepancy of the model to the MIDI interferometric measurements, given the closeness of the model to the mass-loss threshold defined by \citet{mattsson10}. Another critical parameter for dynamic models, in particular for relatively weak winds, is the C/O ratio. After testing a model having a C/O ratio of 1.47, while all the other parameters are identical to the best dynamic model, we found that the model appears, in that case, too much extended compared to the 15\,m baseline VINCI visibility data, whereas the levels of the 60, 30, and 15\,m baselines MIDI visibility data are reproduced relatively well. The transition from windless models to models with considerable mass-loss rates occurs in a very narrow range of stellar parameters especially for the pulsation amplitude, the C/O ratio, and the effective temperature, as already pointed out by \citet{gautschy04}. In summary, a better sampling of parameter space beyond the models currently available \citep{mattsson08,mattsson10}, especially in effective temperature and C/O, could help reduce the discrepancies between models and observations. Such a study is, however, beyond the scope of this paper.\\

(ii) \textbf{The approximation of the stellar pulsation with the sinusoidal piston:}\\
As the star is a semi-regular variable showing two main periods (374 and 1804 days) with harmonics in the smallest period \citep{whitelock97}, using the simple sinusoidal piston to describe the stellar pulsation could be a crude approximation for such a star. The overall wind properties are mostly affected by the pulsation amplitude and not so much by the actual form of the piston motion (see models based on different types of pulsation descriptions presented by \citealt{freytag08}). However, the particular phase coupling between dynamic and radiative properties of the models imposed by the sinusoidal piston might cause some of the discrepancies between models and observations encountered here.\\

(iii) \textbf{The approximation of a complete momentum and position coupling of gas and dust (i.e., absence of drift) and the approximation of the small-particle limit (or Rayleigh limit) in the determination of the dust grain opacity in the dynamic computations:}\\
Gas and dust decouple when the gas of the wind is diluted as is the case for low mass-loss rate stars like \object{R\,Scl}. \citet{sandin03,sandin04} find that decoupling the dust and gas phase increases the dust formation, which could reconcile the dynamic model with the mid-infrared spectro-interferometric measurements. Similar arguments, regarding the difficulty of producing low mass-loss rate models, have also been pointed out by \citet{mattsson10}.\\
It is found by \citet{mattsson10} that dust grains forming in slow wind atmospheres may grow beyond sizes where their opacities can be approximated by the small-particle limit. Moreover, it has been demonstrated by \citet{hoefner08} that radiation pressure on micron-sized silicate grains may play a key role in driving outflows in M-type AGB stars. Similar arguments could also apply to low mass-loss rate carbon-rich stars like \object{R\,Scl}.

\section{Conclusions and perspectives}
\label{conclu-perspec}

In this work, we have presented the first interpretation of combined photometric, spectrometric, and interferometric measurements of a carbon-rich star based on state-of-the-art, self-consistent dynamic atmospheric models.\\
Interferometric observations do not show any significant variability effect at the 16\,m baseline between phases 0.17 and 0.23 in the K band. This is, however, not surprising given the short time interval and because the VINCI broadband measurements average out the short term effects of a shock front in CO or CN lines. No significant variability effect is found for both 15\,m baseline between phases 0.66 and 0.97, and 31\,m baselines between phases 0.90 and 0.97 in the N band. This means that the stellar radiative pressure is not strong enough to reveal a movement of the warm mass dust shells larger than 3 AU, in good agreement with the amplitude of the mass shell of theoretical carbon-rich dynamic models.\\
The spectro-interferometric predictions of the dynamic model atmosphere are in relatively good agreement with the dynamic picture that we deduced from observations. We find rather good agreement between the dynamic model and the spectrophotometric data from 0.4 to 25~$\mu$m. The model agrees relatively well with the time-dependent flux data at 8.5~$\mu$m, whereas the model is too faint compared to the measurements in the 11.3 and 12.5~$\mu$m spectral bands. The 15\,m baseline VINCI visibilities are reproduced well for the two post-maximum brightness phases, meaning that the extension of the dynamic model is suitable in the K-band. In the mid-infrared, the slope of the 60\,m baseline MIDI visibilities in the 8-9~$\mu$m spectral range is reproduced well for all the available phases. This means that the dynamic model has the proper extension to reveal the molecular structures of C$_{2}$H$_{2}$ and HCN located above the stellar photosphere, whereas the non consistent COMARCS+DUSTY model fails to do so.\\
The discrepancy between the dynamic model and the spectro-interferometric data could be related to inaccurate estimation of parameter values of the hydrostatic initial structure, a difference between the parameter values deduced from the hydrostatic modeling and the ones related to the hydrostatic initial structure of the dynamic model, the sinusoidal piston simulating the stellar pulsation, the inclusion of SiC opacities in the a posteriori radiative transfer, or a combination of all these effects. However, it seems that, owing to the strong nonequilibrium process of dust formation in AGB stars, the transition from models without wind to models with considerable mass-loss rates occurs in a very narrow parameter range. The most sensitive parameters allowing this sharp and strong transition between models with and without wind are the effective temperature, the amplitude of the piston velocity, and the C/O ratio. Therefore, it seems necessary to improve the sampling of critical regions in parameter space in the grid of hydrodynamic models for further investigations of the extended structures of low mass-loss carbon stars like \object{R\,Scl}. \\
Finally, the complete dynamic coupling of gas and dust, and the approximation of grain opacities with the small-particle limit in the dynamic calculation, both probably unsuitable for low mass-loss rate object like \object{R\,Scl}, could also contribute to the disagreement between the dynamic model and the spectro-interferometric data. First tests with C-star models based on opacities that take grain size into consideration (Mattsson \& H{\"o}fner, in prep.) show that wind characteristics may be affected considerably in models close to a mass-loss threshold \citep{mattsson10}, such as the one used here to represent the low mass-loss C-star \object{R\,Scl}, resembling recent wind models for M-type AGB stars \citep{hoefner08}.

\begin{acknowledgements}
      
This work is supported by the Austrian Science Fund FWF under the projects P19503-N13, P18939-N16, and P21988-N16. Bernhard Aringer acknowledges funding by the contract ASI-INAF I/016/07/0. Tijl Verhoelst acknowledges support from the  Flemish Fund for Scientific Research (FWO). Authors would like to acknowledge Lars Mattsson for computing and providing the dynamic atmospheric model structures. The authors would like to thank Thorsten Ratzka for helpful discussions on the MIDI reduction software. The authors would like to thank Harald Mutschke and Thomas Posch for providing optical constants of AmC from \citet{rouleau91} in electronic form.

\end{acknowledgements}

\input{13786ref}

\end{document}

%% file: 13786ref.tex
%%%%%%%%%%%%%%%%%%%%%%%% referenc.tex %%%%%%%%%%%%%%%%%%%%%%%%%%%%%%
% sample references
% "physics"
%
% Use this file as a template for your own input.
%
%%%%%%%%%%%%%%%%%%%%%%%% Springer-Verlag %%%%%%%%%%%%%%%%%%%%%%%%%%

%
% BibTeX users please use
% \bibliographystyle{aa}
% \bibliography{}
%
% Non-BibTeX users please use